\documentclass[11pt,a4paper,english]{article}
\usepackage{amsmath,latexsym,amssymb,amsthm,color}
\usepackage{fullpage,times}
\usepackage{ifthen,graphics,epsfig}
\usepackage{graphicx,url,babel,epsfig}                          
\usepackage{amsmath,amsfonts,mathrsfs}
\usepackage{calligra}
\usepackage{float}
\usepackage{diagbox}
\usepackage{xspace}
\usepackage[final]{pdfpages}
\usepackage[normalem]{ulem}

\newcommand{\Xomit}[1]{}
\newcommand{\neighbors}{\mathit{neighbors}}

\newcommand{\CLOCK}{\mathit{CLOCK}}

\newcommand{\df}[1]{}
\newcommand{\ccolor}{{\sc color}}
\newcommand{\colors}{\mathit{colors}}
\newcommand{\tterm}{{\sc term}}
\newcommand{\tokens}{\mathit{tokens}}

\renewcommand{\max}{{\sf max}}
\newcommand{\while}{{\bf while}\xspace}
\newcommand{\col}{{\sf col}}
\renewcommand{\mod}{{\sf ~mod~}}

\newtheorem{theorem}{Theorem}
\newtheorem{lemma}{Lemma}

\newcommand{\toto}{xxx}
\newenvironment{proofT}{\noindent{\bf Proof }} 
{\hspace*{\fill}$\Box_{Theorem~\ref{\toto}}$\par\vspace{3mm}}
\newenvironment{proofL}{\noindent{\bf Proof }} 
{\hspace*{\fill}$\Box_{Lemma~\ref{\toto}}$\par\vspace{3mm}}

\newenvironment{lemma-repeat}[1]{\begin{trivlist}
\item[\hspace{\labelsep}{\bf\noindent Lemma~\ref{#1} }]}%
{\end{trivlist}}

\newenvironment{theorem-repeat}[1]{\begin{trivlist}
\item[\hspace{\labelsep}{\bf\noindent Theorem~\ref{#1} }]}
{\end{trivlist}}

\newcounter{linecounter}
\newcommand{\linenumbering}{\ifthenelse{\value{linecounter}<10}{(0\arabic{linecounter})}{(\arabic{linecounter})}}
\renewcommand{\line}[1]{\refstepcounter{linecounter}\label{#1}\linenumbering}
\newcommand{\resetline}[1]{\setcounter{linecounter}{0}#1}
\renewcommand{\thelinecounter}{\ifnum \value{linecounter} > 9\else 0\fi \arabic{linecounter}}

\newfloat{algorithm}{th}{lop}
\floatname{algorithm}{Algorithm}

\title{\bf  Vertex Coloring  with Communication 
            and Local Memory Constraints 
            in  Synchronous Broadcast  Networks}

\author{ 
        Hicham Lakhlef$^{\ddag}$~~ 
        Michel Raynal$^{\ddag,\star}$~~
        Fran\c{c}ois Ta\"iani$^{\ddag}$~~
~\\~\\
$^{\ddag}$  IRISA, Universit\'e de Rennes,  France \\
$^{\star}$  Institut Universitaire de France\\
{\small{\tt hicham.lakhlef@irisa.fr ~ 
            raynal@irisa.fr ~ 
            francois.taiani@irisa.fr}}
~\\~\\ Tech Report \#2035, 23 pages, April 2016\\ 
       IRISA, University of Rennes 1, France
}
\date{}

\begin{document}

\maketitle

\begin{abstract}
The vertex coloring problem has received a lot of attention 
in the context of synchronous round-based systems where, at each round, 
a process can send a message to all its neighbors, and receive a message
from each of them. Hence, this communication model is 
particularly suited to point-to-point communication channels. 
Several vertex coloring algorithms suited to these systems have been 
proposed. They differ mainly in the  number of rounds they require and 
the number of colors they use. 

This paper considers a broadcast/receive communication model in which
message collisions and message conflicts can occur (a collision occurs when,  
during the same round, messages are sent to the same process by too many 
neighbors; a conflict occurs when a process and one of its neighbors 
broadcast during the same round). This communication model is suited to
systems  where processes share communication bandwidths. More precisely,
the paper considers the case where,  during a round, a process may 
either broadcast a message to its neighbors or receive a message from 
at most $m$ of them. This captures communication-related constraints or 
a local memory constraint stating that,
whatever the number of neighbors of a process, its local memory allows it 
to receive and store at most $m$ messages during each round.  
The paper defines first the  corresponding generic vertex 
multi-coloring problem (a vertex can have several colors). 
It focuses then on tree networks, for which it presents  
a lower bound on the number of colors $K$ that are necessary
(namely, $K=\lceil\frac{\Delta}{m}\rceil+1$, where $\Delta$ is the maximal 
degree of the communication graph),  and an associated  coloring algorithm, 
which is optimal with respect to $K$. 
~ 
\end{abstract}

\thispagestyle{empty}
\newpage
\setcounter{page}{1}

\section{Introduction}

\paragraph{Distributed message-passing synchronous systems}
From a structural point of view, a message-passing system can be
represented by a graph, whose vertices are the processes, and whose
edges are the communication channels.  It is naturally assumed that
the graph is connected.

Differently from asynchronous systems, where there is no notion of
global time accessible to the processes, synchronous message-passing
systems are characterized by upper bounds on message transfer delays
and processing times.  Algorithms for such  systems are
usually designed according to the round-based programming paradigm.
The processes execute a sequence of synchronous rounds, such that,
at every round, each process first sends  a message to its
neighbors, then receives messages from them, and finally executes a
local computation, which depends on its local state and the messages it
has received. The fundamental synchrony property of this model is that
every message is received in the round in which it was sent.  The progress
from one round to the next is a built-in mechanism provided by the
model. Algorithms suited to reliable synchronous systems can be found
in several textbooks (e.g.,~\cite{P00,R13})\footnote{The case where 
processes may exhibit faulty behaviors (such as crashes or Byzantine 
failures) is addressed in several books (e.g.,~\cite{AW04,L96,P00,R10}).}.
When considering reliable synchronous systems, an important issue is
the notion of local algorithm. Those are the algorithms whose 
time complexity (measured by the number of rounds) is smaller than 
the graph diameter~\cite{A81,L92}. 

\paragraph{Distributed graph coloring in point-to-point synchronous systems}
One of the most studied graph problems in the context of  an $n$-process  
reliable synchronous system is the vertex coloring problem, namely any 
process must obtain a color, such that  neighbor processes must have 
different colors (distance-1 coloring),  and the total number of colors 
is reasonably ``small''. 
More generally, the distance-$k$ coloring problem requires that no two 
processes at distance less or equal to $k$, have the same color. 
When considering sequential computing, the optimal distance-1 coloring 
problem is NP-complete~\cite{GJ79}. 

When considering the distance-1 coloring problem in an $n$-process
reliable synchronous system, it has been shown that, if the
communication graph can be logically oriented such that each process
has only one predecessor (e.g., a tree or a ring), $O(\log^*n)$ rounds
are necessary and sufficient to color the processes with at most three
colors~\cite{CV86,L92}\footnote{$\log^* n$ is the number of times the
function $\log$ needs to be iteratively applied in $\log(\log(\log(
...(\log n))))$ to obtain a value $\leq 2$.  As an example, if $n$ 
  is the number of atoms in the universe, $\log^* n \backsimeq 5$.}.
Other distance-1 coloring algorithms are described in several articles
(e.g.~\cite{BE11,BEK14,GPS88,KW06}).  They differ in the number of
rounds they need and in the number of colors they use to implement
distance-1 coloring. Let $\Delta$ be the maximal degree of the graph
(the degree of a vertex is the number of its neighbors).  Both
algorithms in~\cite{BE11,BEK14} color the vertices with $(\Delta+1)$
colors. The first one requires $O(\Delta+\log^* n)$ rounds, while the
second one uses $O(\log \Delta)$ rounds.  An algorithm is described
in~\cite{GPS88} for trees, which uses three colors and $O(\log^* n)$
rounds. Another algorithm presented in the same paper addresses
constant-degree graphs, and uses $(\Delta+1)$ colors and $O(\log^* n)$
rounds. The algorithm presented in~\cite{KW06} requires 
$O(\Delta \log \Delta +\log^* n)$ rounds.  These algorithms assume that 
the processes have distinct identities\footnote{Some initial asymmetry 
is necessary to solve {\it breaking symmetry problems} with a deterministic 
algorithm.},  which define their initial colors. 
They proceed iteratively, each round reducing the total number of colors.
Distributed distance-2 and distance-3 coloring algorithms, suited to 
various synchronous models, are presented 
in~\cite{BM09,BGMBC08,CLSR11,FLR15,GMP02,HT04}.

\paragraph{Motivation  and content of the paper}
The previous reliable synchronous system model assumes that there is a
dedicated (e.g., wired) bi-directional communication channel between
each pair of neighbor processes.  By contrast, this paper considers
a broadcast/receive communication model in which there is no dedicated
communication medium between each pair of neighbor processes.  This
covers practical system deployments, such as wireless networks and
sensor networks.  In such networks, the prevention of collisions
(several neighbors of the same process  broadcast during the
same round), or conflicts (a process and one of its neighbors issue a
broadcast during the same round), does not come for free. In
particular, round-based algorithms that seek to provide deterministic
communication guarantees in these systems must be collision and
conflict-free (\emph{C2}-free in short). 

We are interested in this paper to offer a programming model in which, 
at any round, a process can either broadcast a message to  its neighbors
(conflict-freedom), or receive messages from  at most $m$ of its neighbors 
($m$-collision-freedom).  This means that 
we want to give users a round-based programming abstraction guaranteeing 
conflict-freedom and a weakened form of collision-freedom, that we 
encapsulate under the name C2$m$-freedom (if $m=1$, we have basic C2-freedom).

The ability to simultaneously receive messages from multiple neighbors
can be realized in practice by exploiting multiple frequency
channels\footnote{Depending on the underlying hardware (e.g.,
  multi-frequency bandwidth, duplexer, diplexer), variants of this
  broadcast/receive communication pattern can be envisaged.  The
  algorithms presented in this paper can be modified to take
  them into account.}.  The parameter $m\geq 1$ is motivated by the
following observations.  While a process (e.g., a sensor) may have many
neighbors, it can have constraints on the number of its reception channels, 
or constraints on its  local memory,  that, at each round,
allow it to receive and store messages from only a bounded subset of
its neighbors, namely $m$ of them ($m=1$, gives the classic C2-free model,
while $m\geq \Delta$ assumes no collision can occur as in  the classic
broadcast/receive model
presented previously).  This ``bounded memory'' system parameter can
be seen as a scalability parameter, which allows the degree of a
process (number of its neighbors) to be decoupled from its local
memory size.

C2$m$-freedom can be easily translated as a coloring problem, where
any two neighbors must have different colors (conflict-freedom), and
any process has at most $m$ neighbors with the same color
($m$-collision-freedom).  Once such a coloring is realized, message
consistency is ensured by restricting the processes to broadcast
messages only at the rounds associated with their color. While it is
correct, such a solution can be improved, to allow more
communication to occur during each round.  More precisely, while
guaranteeing C2$m$-freedom, it is possible to allow processes to
broadcast at additional rounds, by allocating multiple colors to
processes. From a graph coloring point of view, this means that,
instead of only one color, a set of colors can be associated with each
process, while obeying the following two constraints: (a) for
any two neighbor processes, the intersection of their color sets
must remain empty; and (b) given any process, no color must appear 
in the color sets of more than $m$ of its neighbors.

We call {\it Coloring with Communication/Memory Constraints} (CCMC) 
the coloring problem described above. More precisely, this problem is denoted 
CCMC$(n,m,K,\geq 1)$, where $n$ is the number of processes (vertices), 
$m$ is the bound on each local memory (bound on the number of 
simultaneous communication from a reception point of view), and
$K$ the maximal number of colors that are allowed. 
``$\geq 1$'' means that there is no constraint on the number of colors that 
that can be assigned to a process. CCMC$(n,m,K,1)$ denotes the problem 
instance where each process is assigned exactly one color. 
From a technical point of view, the paper focuses on tree networks. 
It presents  a lower bound on the value of $K$ for these communication graphs, 
and an algorithm, optimal with respect to $K$, which solves both instances 
of CCMC. 

\paragraph{Roadmap}
The paper is made up of~\ref{sec:conclusion} sections. 
Section~\ref{sec:model} presents the underlying system model. 
Section~\ref{sec:problem}  formally defines the CCMC problem.
Then, considering  tree networks, whose roots are dynamically defined, 
Section~\ref{sec:lower-bound} presents a lower bound on $K$ 
for  CCMC$(n,m,K,1)$ and   CCMC$(n,m,K,\geq 1)$ to be solved.
Section~\ref{sec:algorithm}
presents then  a $K$-optimal algorithm solving  CCMC$(n,m,K,\geq 1)$. 
(from which a solution to CCMC$(n,m,K,1)$ can be easily obtained.)
Section~\ref{sec:proof}  presents a proof of the algorithm.
Finally, Section~\ref{sec:conclusion} concludes the paper.


\section{Synchronous Broadcast/Receive Model}
\label{sec:model}

\paragraph{Processes, initial knowledge, and the communication graph}
The system model consists of $n$ sequential processes denoted $p_1$,
..., $p_n$, connected by a connected communication graph.  When
considering a process $p_i$, $1 \leq i \leq n$, the integer $i$ is called 
its index. Indexes are not known by the processes.  They are only a notation 
convenience used to distinguish processes and their local variables.

Each process $p_i$ has an identity $id_i$, which is known only by itself 
and its neighbors (processes at distance $1$ from it).  The
constant $\mathit{neighbors_i}$ is a local set, known only by $p_i$, 
including the identities of its neighbors (and only them).  
In order for a process $p_i$ not to confuse its
neighbors, it is assumed that no two processes at distance less
than or equal to $2$ have the same identity.  Hence, any two
processes at distance greater than $2$ can have the very same identity.

$\Delta_i$ denotes the degree of process $p_i$ (i.e. $|neighbors_i|$)
and $\Delta$ denotes the maximal degree of the graph
($\max\{\Delta_1,\cdots,\Delta_n\}$).  While each process $p_i$ knows
$\Delta_i$, no process knows $\Delta$ (a process $p_x$ such that
$\Delta_x=\Delta$ does not know that $\Delta_x$ is $\Delta$).

\paragraph{Timing model}
Processing durations are assumed equal to $0$. This is justified by
the following observations: (a) the duration of local computations is
negligible with respect to message transfer delays, and (b) the
processing duration of a message may be considered as a part of its
transfer delay.

Communication is synchronous in the sense that there is an upper bound
$D$ on message transfer delays, and this bound is known by all the
processes (global knowledge). From an algorithm design point of view,
we consider that there is a global clock, denoted $\CLOCK$, which is
increased by $1$, after each period of $D$ physical time units. 
Each value of $\CLOCK$ defines what is usually called a {\it time slot}
or a {\it round}.

\paragraph{Communication operations}
The processes are provided with two operations denoted ${\sf broadcast}()$ 
and ${\sf receive}()$.  A process $p_i$ invokes 
${\sf broadcast}$ {\sc tag}$(m)$ to send the message $m$ (whose type is 
{\sc tag}) to  its neighbors. It is assumed that a process invokes
${\sf broadcast}()$ only at a beginning of a time slot (round).  When a
message {\sc tag}$(m)$ arrives at a process $p_i$, this process is
immediately warned of it, which triggers the execution of the operation
${\sf receive}()$ to obtain and process the message.
Hence, a message is always received 
and processed during the time slot --round-- in which it was broadcast.

From a linguistic point of of view, we use the two following
{\bf when} notations when writing algorithms, where ${\sf predicate}$ 
is a predicate involving $\CLOCK$ and possibly local variables of the
concerned process. 
\vspace{-0.2cm} 
\begin{tabbing} 
$~~~~~~~$
{\bf  when} {\sc  tag}$(m)$ {\bf  is received do} 
                       communication-free processing of the message. \\
$~~~~~~~$
{\bf  when} ${\sf predicate}$ 
                       {\bf do} code entailing at most one ${\sf broadcast}()$
invocation. 
\end{tabbing}

\paragraph{Message collision and message conflict 
           in the $m$-bounded memory model} As announced in the
Introduction, there is no dedicated communication medium for each pair
of communicating processes, and each process has local communication
and memory constraints such that, at every round, it cannot receive
messages from more than $m$ of it neighbors.  If communication is not
controlled, ``message clash'' problems can occur, messages corrupting
each other.  Consider a process $p_i$ these problems are the
following.
\begin{itemize}
\vspace{-0.1cm}
\item 
If more than $m$ neighbors of $p_i$ invoke the operation ${\sf broadcast}()$  
during the same time slot (round), a message {\it collision} occurs.  
\vspace{-0.2cm}
\item 
If $p_i$ and one  of its neighbors invoke ${\sf broadcast}()$  during the 
same time slot (round), a message {\it conflict} occurs.  
\end{itemize}
As indicated in the introduction, an aim of coloring is to prevent 
message clashes from occurring, i.e., in our case, ensures C2$m$-freedom. 
Let us observe that a coloring algorithm must itself be C2$m$-free.


\section{The  Coloring with Communication/Memory Constraints Problem}
\label{sec:problem}

\paragraph{Definition of the CCMC problem}
Let  $\{p_1, \cdots,p_n\}$ be the $n$ vertices of a connected undirected 
graph. As already indicated, $\mathit{neighbors_i}$ denotes the set of the 
neighbors of $p_i$. Let the color domain be the set of non-negative integers, 
and $m$ and $K$ be two positive integers. 
The aim is to associate a set of colors, denoted $\colors_i$, with
each vertex $p_i$, such that the following properties are satisfied.

\begin{itemize}
\vspace{-0.1cm}
\item Conflict-freedom. 
$\forall i,j:$ ($p_i$ and $p_j$ are neighbors) $\Rightarrow$  
 $\colors_i \cap \colors_j = \emptyset$.
\vspace{-0.2cm}
\item $m$-Collision-freedom. 
$\forall i, \forall c: 
|\{ j: p_j \in \mathit{neighbors_i} \wedge c\in \colors_j \}| \leq m$.
\vspace{-0.2cm}
\item Efficiency.  
$| \cup_{1\leq i\leq n} \colors_i | \leq K$.
\end{itemize} 
The first property states the fundamental property of vertex coloring, namely, 
any two neighbors are assigned distinct colors sets. 
The second property states the $m$-constraint coloring  on 
the neighbors of every process, while the third property states an upper 
bound on the total number of colors that can be used. 

As indicated in the Introduction, this problem is denoted 
CCMC($n,m,K,1)$ if each color set is constrained to be a singleton,
and  CCMC($n,m,K,\geq 1)$ if there is no such restriction. 

\begin{figure}[ht]
{\centering\includegraphics[height=6.5cm]{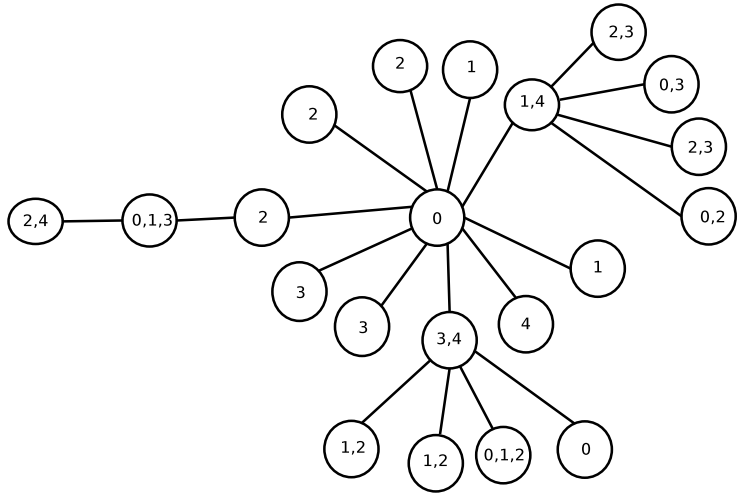} 
\caption{Multi-coloring of a 21-process 10-degree tree 
with the constraint $m=3$ (5 colors)}
\label{exemple-tree-plus-color}
}
\end{figure}

\paragraph{Example}
An example of such a multi-coloring of a 21-process network,
where  $\Delta=10$, and  with the constraint $m=3$, 
is given in Figure~\ref{exemple-tree-plus-color}.
Notice that $K=\lceil \frac{\Delta}{m}\rceil+1=5$
(the color set is $\{0,1,2,3,4\}$).

\paragraph{Particular instances}
The problem instance CCMC$(n,\infty,K,1)$ is nothing other than the 
classical vertex coloring problem, where at most $K$ different colors are 
allowed ($m=\infty$ states that no process imposes a constraint on the colors 
of its neighbors, except that they must be different from its own color).
The problem  instance CCMC$(n,1,K,1)$ is nothing other than the 
classical distance-2 coloring problem (vertices at distance $\leq 2$ 
have different colors). 
 
\paragraph{Using the colors}
The reader can easily see that CCMC($n,m,K,\geq 1$) captures the general 
coloring problem informally stated in the introduction. Once a process $p_i$
has been assigned a set of colors $\colors_i$, at the application programming 
level, it is allowed to broadcast 
a message to neighbors at the rounds (time slots) corresponding to the 
values of $\CLOCK$ such that $(\CLOCK \mod K) \in colors_i$.

\section{CCMC($n,m,K,\geq 1$)  in a Tree Network: Lower Bounds}
\label{sec:lower-bound}

\subsection{An impossibility result}
Considering tree networks, this section presents  a lower bound on $K$:
neither CCMC($n,m,K,1$), nor CCMC($n,m,K,\geq 1$), can be solved for 
$K\leq \lceil \frac{\Delta}{m}\rceil$. The next sections  
will present an algorithm solving CCMC($n,m,K,\geq 1$) in the synchronous 
model described in Section~\ref{sec:model}, and a proof of it. 
As shown next, this algorithm is such $K=\lceil\frac{\Delta}{m}\rceil+1$, 
and is consequently optimal with respect to the total number of colors.

\begin{theorem}
\label{theo:lower-bound}
Neither {\em CCMC($n,m,K,1$)}, nor  {\em CCMC($n,m,K,\geq 1$)} 
can be solved when $K\leq \lceil \frac{\Delta}{m}\rceil$. 
\end{theorem}

\begin{proofT}
Let us first show that  there is no algorithm solving  CCMC($n,m,K,1$) when 
$K\leq \lceil \frac{\Delta}{m}\rceil$. To this end, let us consider a 
process $p_\ell$, which has $\Delta$ neighbors (by the very definition of 
$\Delta$,  there is a such  process). 
Let $\Delta=m\times x +y$, where $0\leq y <m$.  
Hence,  $x=\frac{\Delta-y}{m}=\lfloor\frac{\Delta}{m}\rfloor$ colors 
are needed to color $\Delta-y = m\times x$  processes. 
Moreover, if $y\neq 0$, one more color is needed to color the $y<m$ 
remaining processes. It follows that  $\lceil\frac{\Delta}{m}\rceil$
is a lower bound to color the neighbors of $p_\ell$. 
As $p_\ell$ cannot have the same color as any of its neighbors, it follows that 
at least $\lceil\frac{\Delta}{m}\rceil+1$ are necessary to color
$\{p_i\}\cup \neighbors_i$, which proves the theorem for CCMC($n,m,K,\geq 1$).

Let us observe that an algorithm solving  CCMC($n,m,K,1$) 
can be obtained from an algorithm solving  CCMC($n,m,K,\geq 1$)
by associating with each $p_i$ a single color of its set $colors_i$. 
Hence, any algorithm solving CCMC($n,m,\lceil \frac{\Delta}{m}\rceil,\geq 1$)
can be used to solve  CCMC($n,m,\lceil \frac{\Delta}{m}\rceil,1$).
As  CCMC($n,m,\lceil \frac{\Delta}{m}\rceil,1$) is impossible to solve, 
it follows that CCMC($n,m,\lceil \frac{\Delta}{m}\rceil,\geq 1$) is also 
impossible  to solve. 
\renewcommand{\toto}{theo:lower-bound} 
\end{proofT}

\subsection{A necessary and sufficient condition for multicoloring}
Let CCMC$(n,m,\lceil \frac{\Delta}{m} \rceil +1, >1)$
denote the problem CCMC$(n,m,\lceil \frac{\Delta}{m} \rceil +1, \geq 1)$
where at least one node obtains more than one color. 

\begin{theorem}
\label{theo:multicoloring-condition}
{\em CCMC}$(n,m,\lceil \frac{\Delta}{m} \rceil +1, >1)$
can be solved on a tree of maximal degree $\Delta$, 
if and only if
\vspace{-0.2cm}
$$\exists i : \lceil \frac{\Delta}{m} \rceil +1 > \max
\left(\; \left\{ \left\lceil \textstyle\frac{\Delta_i}{m}\right\rceil \right\} 
\cup \left\{ \left\lfloor \left. \textstyle\frac{\Delta_j}{m}\right\rfloor \;
\right|\; p_j \in \neighbors_i \right\}\;\right) + 1.$$
\end{theorem} 

\noindent
The proof of this theorem appears in Appendix~\ref{proof:theo-multicoloring}.

\section{CCMC($n,m,K,\geq 1$)  in a Tree Network: Algorithm}
\label{sec:algorithm}
 The algorithm presented in this section use as a skeleton  a parallel
traversal of a tree~\cite{R13}. Such a traversal is implemented by
control messages that visit all the processes, followed by a control
flow that returns at the process that launched the tree traversal.

Algorithm~\ref{fig:tree-multicoloring-algorithm} is a C2$m$-free 
algorithm that solves the 
CCMC($n,m,\lceil \frac{\Delta}{m}\rceil,\geq 1$) problem. 
It assumes that a single process initially receives an external
message {\sc start}$()$, which dynamically defines it as the root of the tree.
This message and the fact that processes at distance smaller or equal to $2$ 
do not have the same identity provide  the initial asymmetry from which a 
deterministic coloring algorithm can be built. 
The reception of the message {\sc start}$()$ causes the receiving process 
(say $p_r$) to simulate the reception of a fictitious message 
{\sc color}$()$, which initiates the sequential traversal.

\paragraph{Messages}
The algorithm uses two types of messages, denoted {\sc color}$()$
and {\sc term}$()$.
\begin{itemize}
\vspace{-0.2cm}
\item
The messages {\sc color}$()$ implement a control flow visiting in parallel 
the processes of the tree from the root to the leaves.  Each of them carries 
three values, denoted $sender$,  $cl\_map$, and $max\_cl$. 
\begin{itemize}
\vspace{-0.1cm}
\item $sender$ is the identity of the sender of the message. 
If it is the first message \ccolor$()$ received by a process $p_i$, 
$sender$ defines the  parent of $p_i$ in the tree. 
\vspace{-0.1cm}
\item $cl\_map$ is a dictionary data structure with one entry for 
each element in $\neighbors_x \cup \{id_x\}$, where $p_x$ is the 
sender of the message  \ccolor$()$. 
 $cl\_map[id_x]$ is the set of colors currently assigned to 
the sender and, for  each  $id_j \in  neighbor_x$,  
$cl\_map[id_j]$ is the set of colors that $p_x$ proposes for $p_j$. 
\vspace{-0.1cm}
\item  
$max\_cl$ is an integer defining the color domain used by the sender, 
namely  the color set  $\{0,1,\ldots, (max\_cl-1)\}$.
Each child $p_i$ of the message sender will use the color domain 
defined by ${\sf max}(max\_cl,\sigma_i)$ to propose colors to its own children
($\sigma_i$ is defined below). Moreover, all the children of the sender 
will use the same slot span $\{0,1,\ldots, (max\_cl-1)\}$
to broadcast their messages. This  ensures that their message  
broadcasts will be collision-free\footnote{ As we will see, conflicts are 
prevented by the message exchange pattern imposed by the algorithm.}. 
\end{itemize}
\vspace{-0.2cm}
\item 
The messages {\sc term}$()$ propagate the return of the control flow
from the leaves to the root.  Each message {\sc term}$()$ carries two
values: the identity of the destination process (as this message is broadcast, 
this allows any receiver to know if the message is for it), and the identity 
of the sender.
\end{itemize}

\paragraph{Local variables}
Each process $p_i$ manages the following local variables. 
The constant $\Delta_i=|\neighbors_i|$ is the degree of $p_i$, while the 
constant  $\sigma_i=\lceil\frac{\Delta_i}{m}\rceil +1$ is the number
of colors needed to color the star graph made up of $p_i$ and its neighbors. 
\begin{itemize}
\vspace{-0.1cm}
\item $state_i$ (initialized to $0$) is used by $p_i$ to manage the
  progress of the tree traversal.  Each process traverses five
  different states during the execution of the algorithm. States $1$
  and $3$ are active states: a process in state $1$ broadcasts a
  \ccolor$()$ message for its neighbors, while a process in state $3$
  broadcasts a message \tterm$()$ which has a meaning only for its parent.  
  States $0$ and $2$ are waiting states in which a  
  process listens on the broadcast channels but  cannot send any message. 
  Finally, state $4$ identifies local  termination.
\vspace{-0.2cm}
\item $parent_i$ stores the identity of the process $p_j$ from which
  $p_i$ receives a message {\sc color}$()$ for the first time
  (hence $p_j$ is the parent of $p_i$ in the tree).  The root $p_r$ of
  the tree, defined by the reception of the external message {\sc start}$()$, 
  is the only process such that $parent_r = id_r$.
\vspace{-0.2cm}
\item $colored_i$ is a set containing the identities of the neighbors
  of $p_i$ that have been colored.
\vspace{-0.2cm}
\item $to\_color_i$ is the set of neighbors to which $p_i$ must 
propagate the coloring (network traversal).
\vspace{-0.2cm}
\item $color\_map_i[\neighbors_i\cup \{id_i\}]$ 
is a dictionary data structure where $p_i$ stores colors 
of its neighbors in  $color\_map_i[\neighbors_i]$, 
and its own colors in  $color\_map_i[id_i]$;  
$\colors_i$ is used as a synonym of  $color\_map_i[id_i]$. 
\vspace{-0.2cm}
\item $max\_cl_i$  defines both the color domain from which
$p_i$ can color its children, and the time slots (rounds)  at which  
its children  will be allowed to broadcast. 
\vspace{-0.2cm}
\item $slot\_span_i$ is set to the value $max\_cl$ 
carried by the message \ccolor$()$ received by $p_i$ from its parent. 
As this value is the same for all the children of its parent, 
they will use the same slot span to define the slots during 
which each child  will be allowed  to broadcast messages. 
\end{itemize}

\begin{algorithm}[h!]
\centering{
\fbox{
\begin{minipage}[t]{150mm}
\footnotesize
\renewcommand{\baselinestretch}{2.5}
\resetline
\begin{tabbing}
aaaA\=aaaA\=aaA\=aaA\=aaA\=aaA\kill

{\bf  Initialization:} $\sigma_i= \lceil\frac{\Delta_i}{m}\rceil +1$; 
      $state_i\leftarrow 0$; $\colors_i \leftarrow \emptyset$;
      $\colors_i$ is a synonym of $color\_map_i[id_i]$.
\\~\\

\line{MC-01}  \> 
{\bf  w}\={\bf hen} {\sc  start}$()$ {\bf  is received do} 
      \%  a single process $p_i$ receives this external message  \% \\

\line{MC-02}  \> \> 
$p_i$ executes lines~\ref{MC-04}-\ref{MC-25} as if it received the message 
{\sc color}$(id_i, cl\_map, \sigma_i)$ \\
\>\> where 
     $cl\_map \left[ id_i\right]  = \{ (\CLOCK+1) {\sf ~mod~}\sigma_i \}$ .\\~\\

\line{MC-03} \> 
{\bf  when} 
{\sc color}$(sender, cl\_map, max\_cl)$ {\bf  is received} {\bf do} \\

\line{MC-04} \>\> 
{\bf if}  (first message {\sc color}$()$ received) \\

\line{MC-05} \>\>  {\bf then}  \=        
  $parent_i \leftarrow sender$; 
  $color\_map_i[parent_i] \leftarrow cl\_map[sender]$; \\
    
\line{MC-06}  \>\> \>  
$colored_i   \leftarrow \{sender\}$;
$to\_color_i \leftarrow \mathit{neighbors_i}\setminus\{sender\} $;\\

\line{MC-07}  \>\> \> 
 $color\_map_i[id_i] \leftarrow cl\_map[id_i]$; 
  ~~~\% Synonym of $\colors_i$  \% \\
    
\line{MC-08}
\>\> \> $max\_cl_i \leftarrow {\sf max}(max\_cl,\sigma_i)$;
     $slot\_span_i \leftarrow max\_cl$; \\   
 
\line{MC-09} \>\> \> {\bf if} \= $(to\_color_i \neq  \emptyset)$  

 ~~~~ \%  next lines: $\tokens_i$ is a multiset \%  \\

\line{MC-10}  \>\>\>\> 

    \= {\bf then} \= $\tokens_i \leftarrow$ \= ~~~ \=
$\{$ $m$ tokens with color $x$, 
  for each $x \in \big([0..(max\_cl_i-1)] \setminus \colors_i\big) ~\}$\\

\>\>\>\>\>\>\> 
$\setminus$ 
\> $\{$~one token with color $z$, for each $z \in color\_map_i[parent_i]~\}$;\\

\line{MC-11}  \>\>\>\>\>\> 
{\bf while} \= $(|\tokens_i|< |to\_colors_i|)$ {\bf do}\\

\line{MC-12}   \>\>\>\>\>\>\> 
{\bf if} $(|colors_i|>1)$ \= {\bf then} \=
     {\bf let} $cl \in \colors_i$; suppress $cl$ from $\colors_i$\\

\line{MC-13}   \>\>\>\>\>\>\>\>\> 
    add $m$ tokens colored $cl$ to $\tokens_i$\\

\line{MC-14}   \>\>\>\>\>\>\>\> {\bf else} \>
    {\bf let}   $cl$ be the maximal color in $color\_map_i[parent_i]$;  \\

\line{MC-15}   \>\>\>\>\>\>\>\>\> 
       add one  token colored $cl$ to $\tokens_i$;\\

\line{MC-16}   \>\>\>\>\>\>\>\>\> 
$color\_map_i[parent_i] \leftarrow color\_map_i[parent_i]\setminus \{ cl\}$\\

\line{MC-17} \>\>\>\>\>\>\> {\bf end if}\\

\line{MC-18} \>\>\>\>\>\> {\bf end while};\\

\line{MC-19}  \>\>\>\>\> \> 
Extract $|to\_colors_i|$ non-empty non-intersecting multisets $tk[id]$ 
(where $id\in to\_color_i$)    \\
\>\>\>\>\> \> 
from $\tokens_i$ 
such that no  $tk[id]$ contains several tokens with the the same color;\\

\line{MC-20}  \>\>\>\>\> \> 
{\bf for each} $id \in to\_color_i$ {\bf do}
$color\_map_i[id] \leftarrow \{$colors of the tokens in $tk[id]\}$  
{\bf end for};\\

\line{MC-21} \>\>\>\>\>\> 
 $state_i \leftarrow 1$ ~~ \% $p_i$ has children \% \\

\line{MC-22}  \>\>\> \> 
  \> {\bf else} \>  $state_i \leftarrow 3$  ~~ \% $p_i$ is a leaf \% \\

\line{MC-23} \>\> \> {\bf end if}\\

\line{MC-24}  \>\> {\bf else} 
$color\_map_i[id_i] \leftarrow color\_map_i[id_i] \cap cl\_map[id_i]$\\

\line{MC-25}  \>\>  {\bf end if}.\\~\\

\line{MC-26}  \> {\bf when} 
$\big( (\CLOCK {\sf ~mod~} slot\_span_i)~\in colors_i) 
                               \wedge (state_i\in \{1,3\})\big)$ {\bf do}\\

\line{MC-27}   \>\>
 {\bf case} \=  $(state_i=1)$  {\bf then} 
 ${\sf broadcast}$  {\sc color}$(id_i,color\_map_i, max\_cl_i)$; 
     $state_i\leftarrow 2$ \\

\line{MC-28}  \>\>\>  $(state_i=3)$  {\bf then} 

 ${\sf broadcast}$ {\sc term}$(parent_i,id_i)$; 
 $state_i\leftarrow 4$   \% $p_i$'s subtree is colored \% \\

\line{MC-29} \>\>  {\bf end case}. \\~\\

\line{MC-30}  \>
{\bf when}  {\sc term}$(dest,id)$ {\bf is received} {\bf do} \\

\line{MC-31}  \>\> {\bf if} $(dest \neq id_i)$ 
   {\bf then} discard the message
      (do not execute lines~\ref{MC-25}-\ref{MC-28}) {\bf end if};\\

\line{MC-32}  \>\> $colored_i  \leftarrow colored_i  \cup \{ id\}$;  \\

\line{MC-33}    \>\> {\bf if} \= ($colored_i= \neighbors_i$)  \\

\line{MC-34}   \>\>\> {\bf then if}  ($parent_i=id_i$) 
 {\bf then}  the root $p_i$ claims termination 
 {\bf else}   $state_i\leftarrow 3$  {\bf end if} \\

\line{MC-35}  \>\> {\bf end if}.

\end{tabbing}
\normalsize
\end{minipage}
}
\caption{C2$m$-free algorithm solving  
CCMC($n,m,\lceil \frac{\Delta}{m}\rceil+1,\geq 1$) in tree networks 
(code for $p_i$)}
\label{fig:tree-multicoloring-algorithm}
}
\end{algorithm}

\paragraph{Initial state}
In its initial state ($state_i=0$), a process $p_i$ waits for a message 
\ccolor$()$. As already indicated, a single process receives the external 
message {\sc start}$()$, which defines it at the root process. 
It is assumed that $\CLOCK=0$ when a process receives this message. 
When it receives it,  the corresponding process $p_i$ simulates the
reception of the message {\sc color}$(id_i,cl\_map,\sigma_i)$ 
where  $cl\_map[id_i]$ defines its color,   namely, 
$(\CLOCK+1) {\sf  ~mod~} \sigma_i$ (lines~\ref{MC-01}-\ref{MC-02}). 
Hence, at round number $1$, the root will send a message \ccolor$()$ 
to its children (lines~\ref{MC-19}-\ref{MC-20}).

\paragraph{Algorithm: reception of a message {\sc color}$()$}
When a process $p_i$ receives a message \ccolor$()$ for the first time, it is 
visited by the network traversal, and must consequently (a) obtain an initial
color set, and (b) propagate the the network traversal, if it has children.
The  processing by $p_i$ of this first message 
\ccolor$(sender,cl\_map,max\_cl)$  is done at 
lines~\ref{MC-05}-\ref{MC-23}.   
First, $p_i$ saves the identity of its parent (the sender of the message)
and its proposed color set (line~\ref{MC-05}), initializes $colored_i$
to $\{sender\}$, and  $to\_color_i$ to its other neighbors (line~\ref{MC-06}).  
Then $p_i$ obtains a color set proposal 
from the dictionary $cl\_map$ carried by the 
message (line~\ref{MC-07}),  computes the value $max\_cl_i$ from which its 
color palette will be defined, and saves the value $max\_cl$ carried by the 
message \ccolor$()$ in the local variable $slot\_span_i$ (line~\ref{MC-08}).
Let us remind that the value $max\_cl_i$ allows it to know the color domain 
used up to now, and the rounds at which it will be able to broadcast messages
(during the execution of the algorithm) in a collision-free way.  

Then, the behavior of $p_i$ depends on the value of $to\_color_i$.
If  $to\_color_i$ is empty, $p_i$ is a leaf, and there is no more process 
to color from it. Hence, $p_i$  proceeds to state $3$ (line~\ref{MC-22}). 

If  $to\_color_i$ is not empty, $p_i$ has children. It has consequently to 
propose a set of colors for each of them, and save these proposals in its
local dictionary  $color\_map_i[\neighbors_i]$.  
To this end,  $p_i$ computes first the domain of colors it can use, namely, 
the set  $\{0,1,\ldots,(max\_cl_i-1)\}$, and considers that each of these 
colors $c$ is represented by $m$  tokens colored $c$.  Then, it computes 
the multiset\footnote{Differently from a set, a {\it multiset} (also
called a {\it  a bag}), can contain several times the same element.
Hence, while $\{a,b,c\}$ and $\{a,b,a,c,c,c\}$ are the same set, 
they are different multisets.}, denoted $\tokens_i$, containing all the  
colored tokens it can use to build a color set proposal 
for each of its children (line~\ref{MC-10}). 
The multiset  $\tokens_i$ is initially made up of  all possible 
colored tokens, from which are suppressed (a) all
tokens associated with the colors of $p_i$ itself, and, (b)
one colored token for each color in $color\_map_i[parent_i]$ 
(this is because, from a coloring point of view,  its parent was allocated 
one such colored token for each of its colors).  

Then, $p_i$ checks if it has enough colored tokens to allocate at least one 
colored token to each of its children (assigning thereby the color of the 
token to the corresponding child). 
If the predicate $|\tokens_i|\geq|to\_color_i|$ is satisfied, $p_i$ has 
enough colored tokens and can proceed to assign set of colors to its children
(lines~\ref{MC-19}-\ref{MC-20}). Differently, if  the predicate 
$|\tokens_i|<|to\_color_i|$ is satisfied, $p_i$  has more children than 
colored  tokens. Hence, it must find more colored tokens. 
For that,  if $\colors_i$ (i.e., $color\_map_i[id_i]$) has more than one  color,
$p_i$ suppresses one color from $\colors_i$, adds the $m$ associated
colored tokens to the multiset $\tokens_i$ (lines~\ref{MC-12}-\ref{MC-13}), 
and re-enters the ``while'' loop (line~\ref{MC-11}). 
If $\colors_i$  has a single  color, this color cannot be suppressed from
$\colors_i$. In this case, $p_i$ considers the color set of its parent
($color\_map_i[parent_i]$), takes the maximal color of this set,
suppresses it from  $color\_map_i[parent_i]$, adds the associated
colored token to the multiset $\tokens_i$, and --as before-- re-enters 
the ``while'' loop (line~\ref{MC-15}). Only one  token
colored $cl$ is available  because the  $(m-1)$ other tokens  
colored $cl$ were already added into the multiset $\tokens_i$ during its 
initialization at line~\ref{MC-10}.

As already said, when the  predicate $|\tokens_i|<|to\_color_i|$ 
(line~\ref{MC-11}) 
becomes false,  $\tokens_i$ contains enough colored tokens to assign 
 to its children. This assignment is done at lines~\ref{MC-19}-\ref{MC-20}. 
Let $ch=|to\_color_i|$ (number of children of $p_i$); 
$p_i$ extracts $ch$ pairwise disjoint and non-empty subsets of the multiset 
$\tokens_i$, and assigns each of them to a different neighbor. 
``Non-empty non-intersecting multisets'' used at line~\ref{MC-19} 
means that, if each of $z$ multisets $tk[id_x]$ contains a token with the 
same color, this colored token appears at least $z$ times in the multiset 
$\tokens_i$.

If the message \ccolor$(sender,cl\_map,-)$ received by $p_i$ is not 
the first one, it was sent by one of its  children. In this case, 
$p_i$ keeps in  its color set $color\_map_i[id_i]$ ($\colors_i$)
only colors allowed  by its child $sender$ (line~\ref{MC-24}). 
Hence, when $p_i$ has received a message  \ccolor$()$ from each of 
its children, its color set $\colors_i$ has its final value.

\paragraph{Algorithm: broadcast of a message}
A process $p_i$ is allowed to broadcast a message only at the rounds
corresponding to a color it obtained (a color in 
$\colors_i = color\_map_i[id_i]$ computed at lines~\ref{MC-07},
\ref{MC-12}, and~\ref{MC-24}), provided that its current 
local state is $1$ or $3$  (line~\ref{MC-26}). 

If $state_i=1$, $p_i$ received previously a message {\sc color}$()$, 
which entailed its initial coloring and a proposal to color its children 
(lines~\ref{MC-09}-\ref{MC-21}). In this case, $p_i$  propagates the tree
traversal by broadcasting a message {\sc color}$()$ (line~\ref{MC-27}), 
which will provide each of its children with a coloring  proposal. 
Process $p_i$ then progresses to  the local  waiting state $2$. 

If $state_i=3$, the coloring of the tree rooted at $p_i$ is terminated. 
Process $p_i$ consequently broadcasts the message {\sc term}$(parent_i,id_i)$ 
to inform its parent of it. It also progresses from state $3$ to state $4$,
which indicates its local termination (line~\ref{MC-28}).

\paragraph{Algorithm: reception of a message {\sc term}$()$}
When a process $p_i$ receives such a message it discards it if it is not 
the intended destination process (line~\ref{MC-31}). If the message is for it, 
$p_i$ adds the sender identity to the set $ colored_i$  (line~\ref{MC-32}). 
Finally, if  $colored_i=\neighbors_i$, $p_i$ learns that the subtree rooted 
at it is colored (line~\ref{MC-33}).  It follows that, if $p_i$ is the root 
($parent_i=i$), it learns that the algorithm terminated. Otherwise, it enters 
state $3$, that will  direct it to report to its parent the termination of 
the coloring of the  subtree rooted at it. 

\paragraph{Solving CCMC($n,m,K,1$)  in a tree}
Algorithm~\ref{fig:tree-multicoloring-algorithm} can be easily modified 
to solve  CCMC($n,m,K,1$). When a process enters  state $3$
(at line~\ref{MC-22} or line~\ref{MC-34}), 
it reduces $color\_map_i[id_i]$ (i.e., $\colors_i$) to obtain a singleton.

\section{CCMC($n,m,K,\geq 1$)  in a Tree Network: Cost and Proof}
\label{sec:proof}
\label{sec:algo-proof}

The proof assumes $n>1$. 
Let us remember that $\colors_i$ and $color\_map_i[id_i]$ are 
the same local variable of $p_i$, and $p_r$ denotes the dynamically 
defined root process.  

\paragraph{Cost of the algorithm}  
Each non-leaf process broadcasts one
message {\sc color}$()$, and each non-root process broadcasts one 
message {\sc term}$()$. Let $x$ be the number of leaves. There are
consequently $(2n-(x+1))$ broadcasts. As $\Delta \leq x+1$ (\footnote{Let $p_i$ be the process that has $\Delta$ as degree. 
If  $p_i$ is the root of the tree, the tree contains at least $\Delta$ leaf processes. 
This is because each neighbor of $p_i$ is either a leaf or the root of a subtree that has at least one leaf process. 
And if $p_i$ is not the root of the tree, $p_i$ possesses $\Delta-1$ children, and the number of leaf processes is at least $\Delta-1$ following a similar reasoning.}), the number
of broadcast is upper bounded by $2n-\Delta$.

Given an execution whose  dynamically defined root is the process $p_r$, 
let $d$ be the height of the corresponding tree. 
The root computes the colors defining the slots (rounds) at which its children 
can broadcast the messages \ccolor$()$ and {\sc term}$()$.  
These colors span the interval  $[0..\lceil\frac{\Delta_r}{m}\rceil]$, 
which means that the broadcasts of messages {\sc color}$()$ by the
processes at the first  level of the tree span at most
$\lceil\frac{\Delta_r}{m}\rceil+1$ rounds. The same broadcast pattern 
occurs at each level of the tree. It follows that the visit of the tree by 
the messages \ccolor$()$ requires at most $d\lceil\frac{\Delta}{m}\rceil$ 
rounds. 
As the same occurs for the the messages  {\sc term}$()$, returning from 
the leaves to the root, it follows that the time complexity of the algorithm
is $O(d\lceil\frac{\Delta}{m}\rceil)$.

\begin{lemma}
\label{lemma:noconflict}
\label{lemma:conflict-freedom}
Algorithm~{\em\ref{fig:tree-multicoloring-algorithm}} is conflict-free.
\end{lemma}

\begin{proofL}
The algorithm uses two types of messages: {\sc color}$()$ and {\sc term}$()$.  
We first show conflict-freedom for {\sc color}$()$
messages (if a process broadcasts a message  {\sc color}$()$, 
none of its neighbors is broadcasting any message in the same  round).  
Let us first notice that a process $p_i$
broadcasts at most one message {\sc color}$()$, and one message 
{\sc term}$()$ (this is due to the guard $state_i\in\{1,3\}$,
line~\ref{MC-26}, and the fact that the broadcast of a message makes
its sender progress to the waiting state $2$ or $4$). Moreover, let us
make the following observations.
\begin{itemize}
\vspace{-0.2cm}
\item
Observation 1: 
The first message sent by any node is of type {\sc color}$()$ 
(line~\ref{MC-27}).
\vspace{-0.2cm}
\item
Observation 2: Except for the root process, a message  {\sc color}$()$ 
is always broadcast by a process after it received a message {\sc color}$()$ 
(which triggers the execution of lines~\ref{MC-03}-\ref{MC-25}).
\vspace{-0.2cm}
\item
Observation 3: Except for leaf processes, a message  {\sc term}$()$ is 
always broadcast by a process after it received a message {\sc term}$()$  from 
each of its children (lines~\ref{MC-30}-\ref{MC-35} and line~\ref{MC-28}.).

\end{itemize}
Observations 1 and 2  imply that when the root process broadcasts its 
{\sc color}$()$ message, none of its neighbors is broadcasting a message, 
and they all receive the root's {\sc color}$()$ message without conflict.
Let us now consider a process $p_i$, different from the root, 
which receives its first message {\sc  color}$_k()$ (from its parent $p_k$). 
Because there is no cycle in the communication graph (a tree), all
the children of $p_i$ ($\mathit{neighbors}_i \setminus \{p_k\}$) are
in state $0$, waiting for their  {\sc color}$()$ message. 
Moreover, due to Observations 1 and 2, they will receive from $p_i$ 
their  message {\sc color}$()$  without conflict. 
After sending its {\sc color}$()$ message, $p_i$'s parent
$p_k$ remains in the waiting state $2$ until it receives a {\sc term}$()$
message from all its children (lines~\ref{MC-32}-\ref{MC-33}), which
include $p_i$. As a consequence, $p_k$ is not broadcasting any message
in the round in which it receives $p_i$'s {\sc color}$()$ message, 
which is consequently received without conflict by all its neighbors.

As far the messages {\sc term}$()$ are concerned we have the following. 
Initially, only a leaf process can broadcast a message, and when it does it, 
its parent is in the waiting state $2$  (since it broadcast a message
{\sc color}$()$ at line~\ref{MC-27} and it must receive  messages 
{\sc term}$()$ to proceed to state $3$).  Hence a message   {\sc term}$()$
broadcast by a leaf cannot entail conflict. Let us now consider a
non-leaf process $p_i$. It follows from  Observation 3
that $p_i$  can broadcast a message {\sc term}$()$ only when its children 
are in state $4$ (in which they cannot broadcast),  and its parent 
(because it has not yet received a message  {\sc term}$()$ from each of 
its children) is in the  waiting state $2$. Hence, we conclude that 
the broadcast of a message  {\sc term}$()$ by a non-leaf process is 
conflict-free, which concludes the proof of the lemma. 
\renewcommand{\toto}{lemma:noconflict}
\end{proofL}

\paragraph{Definition}
A message \ccolor$(sender,cl\_map,max\_cl)$ is {\it well-formed} if 
its content satisfies the following properties. Let $sender=id_i$.  
\begin{itemize}
\vspace{-0.2cm}
\item[M1]  The keys of the dictionary data structure $cl\_map$
are the identities in $neighbors_i \cup \{id_i\}$. 
\vspace{-0.2cm}
\item[M2] 
$\forall ~id\in (\neighbors_i \cup \{id_i\}):~ cl\_map[id]\neq \emptyset$.
\vspace{-0.2cm}
\item[M3] 
 $\forall ~id\in \neighbors_i:~ cl\_map[id]\cap cl\_map[id_i]= \emptyset$.
\vspace{-0.2cm}
\item[M4]  $\forall c:~ 
    |\{j: (id_j\in \neighbors_i)\wedge(c\in cl\_map[id_j])\}| \leq m$.
\vspace{-0.2cm} 
\item[M5]  $1< max\_cl \leq \lceil\frac{\Delta}{m}\rceil+1$. 
\vspace{-0.2cm} 
\item[M6] $\forall ~id\in (\neighbors_i \cup \{id_i\}):~ 
       cl\_map[id] \subseteq  [0..max\_cl-1]$.
\end{itemize}

Once established in Lemma~\ref{lemma:well-formed-message}, not
all properties M1-M6 will be explicitly used in the lemmas
that follow.  They are used by induction to proceed from one 
well-formed message to another one.

\begin{lemma}
\label{lemma:while-loop-termination}
If a message \ccolor$(sender,cl\_map,max\_cl)$ 
received by a process $p_i\neq p_r$ 
is  well-formed and entails the execution 
of lines~{\em{\ref{MC-05}-\ref{MC-23}}},
the {\em\textbf{while}} loop (lines~{\em{\ref{MC-11}-\ref{MC-18}}})
terminates, and,  when $p_i$ exits the loop, the sets $\colors_i$ and 
$color\_map_i[parent_i]$ are not empty, and their intersection is empty. 
\end{lemma}

\begin{proofL}
Let us consider a process $p_i \neq p_r$ that receives a
well-formed \ccolor$_j(sender,cl\_map,max\_cl)$ message 
from $p_j$. 
Let us assume \ccolor$()$ causes $p_i$ to start executing the 
lines~\ref{MC-05}-\ref{MC-23}, i.e., \ccolor$()$
is the first such message received by $p_i$. The body of
the \while loop contains two lines (lines~\ref{MC-12}
and~\ref{MC-14}) that select elements from two sets, $\colors_i$ and
$color\_map_i[parent_i]$ respectively. 

Before discussing the  termination of the \while loop, we  show that 
lines~\ref{MC-12} and~\ref{MC-14}
are \emph{well-defined}, i.e. the sets from which the elements
are selected are non-empty. To this aim, we prove by
induction that the following invariant holds in each iteration of the loop:
\begin{align}
    color\_map_i[parent_i] &\neq \emptyset,\label{eq:colormapparentnonempty} \\
    \colors_i &\neq \emptyset, \label{eq:colorinonempty} \\
    |\tokens_i| &= m \times max\_cl_i - 
    m \times |\colors_i| - |color\_map_i[parent_i]|. \label{eq:tokencard}
\end{align}
Just before the loop (i.e., before line~\ref{MC-11}), Assertion 
(\ref{eq:colormapparentnonempty}) follows from the assignment to
$color\_map_i[parent_i]$ at line~\ref{MC-05} and the property M2 of
\ccolor$_j$() ($id_j=parent_i$).  
Assertion  (\ref{eq:colorinonempty}) also follows from M2
($colors_i$ is synonym of $color\_map_i[id_i]$). 
Assertion (\ref{eq:tokencard}) follows from M3, M6, and the
initialization of $max\_cl_i$ at line~\ref{MC-08}.

Let us now assume that  Assertion (\ref{eq:colormapparentnonempty}) 
holds at the start of a loop iteration (i.e., just before
lines~\ref{MC-12}). There are two cases.
\begin{itemize}
\vspace{-0.2cm}
\item If $|\colors_i|>1$, lines~\ref{MC-14}-\ref{MC-16} are not
executed, and consequently $color\_map_i[parent_i]$ is not modified.
It follows from the induction assumption that Assertion 
(\ref{eq:colormapparentnonempty}) still holds.
\vspace{-0.2cm}
\item If $|\colors_i|\leq1$, we have the following. 
Because we are in the {\bf while} loop, we have $|\tokens_i|< |to\_colors_i|$, 
which, combined with Assertion (\ref{eq:tokencard}), implies\\
\centerline{
    $|to\_colors_i| > m \times max\_cl_i - m \times 
                             |\colors_i| - |color\_map_i[parent_i]|$,}\\
from which we derive\\
~\\ \vspace{-1.4cm}
\begin{tabbing}
aA\=aaaA\=aaA\=aaA\=aaA\=aaA\kill
\>    $|color\_map_i[parent_i]|$~~ \= $>~ m \times max\_cl_i - 
                                  m \times |\colors_i| - |to\_colors_i|$,\\ 
\> \> $>~ m \times max\_cl_i - m \times |\colors_i| - (\Delta_i -1)$
                                 ~~~~~ \= (because of line~\ref{MC-06}),\\
\> \> $>~ m \times \sigma_i - m \times |\colors_i| - (\Delta_i -1)$ 
                                           \> (because of line~\ref{MC-08}),\\
\> \> $>~ m \times (\lceil\frac{\Delta_i}{m}\rceil +1) - m \times 
                     |\colors_i| - (\Delta_i -1)$  \> (by definition),\\   
\> \> $>~ \Delta_i +m - m \times |\colors_i| - (\Delta_i -1)$ \>(arithmetic),\\
\> \> $>~ m \times (1-|\colors_i|) + 1$.

\end{tabbing} 
\vspace{-0.1cm}
Hence, because $|\colors_i|\le 1$, we obtain   
$|color\_map_i[parent_i]| > 1$, which means that $p_i$'s local variable 
$color\_map_i[parent_i]$ contains at least two elements before the execution
of line~\ref{MC-12}. Because only one color 
is removed from $color\_map_i[parent_i]$, this local variable 
remains non-empty after line~\ref{MC-16}, thus proving Assertion 
(\ref{eq:colormapparentnonempty}).
\end{itemize}
\vspace{-0.2cm}
Let us now  assume that both  Assertion (\ref{eq:colorinonempty}) and 
 Assertion (\ref{eq:tokencard}) hold at the start of a loop iteration 
(i.e., just before line~\ref{MC-12}). There are two cases. 
\begin{itemize}
\vspace{-0.2cm}
\item
Case $|\colors_i|>1$. In this case we have:
\emph{(i)} one color is removed from $\colors_i$,
\emph{(ii)} $m$ colored tokens are added to $\tokens_i$, and \emph{(iii)}
 $color\_map_i[parent_i]$ remains unchanged.  $|\colors_i|>1$ and
\emph{(i)} imply that  Assertion (\ref{eq:colorinonempty}) remains true; and
\emph{(i)} and \emph{(ii)} mean that  Assertion (\ref{eq:tokencard}) is 
preserved.
\vspace{-0.2cm}
\item
Case  $|\colors_i|\le 1$. In this case we have:  \emph{(i)} one color is 
removed from $color\_map_i[parent_i]$, and one colored token added 
to $\tokens_i$, and
\emph{(ii)} $\colors_i$ stays unchanged. \emph{(i)} implies that
 Assertion (\ref{eq:tokencard}) remains true, and \emph{(ii)} ensures
 Assertion (\ref{eq:colorinonempty}) by assumption.
\end{itemize}
\vspace{-0.2cm}
This concludes the proof that the three
assertions~(\ref{eq:colormapparentnonempty})--(\ref{eq:tokencard}) are
a loop invariant. Hence,  Assertion 
(\ref{eq:colormapparentnonempty}) and Assertion (\ref{eq:colorinonempty})
imply that lines~\ref{MC-12} and~\ref{MC-14} are well-defined.

Let us now observe that, in each iteration of the loop, new colored
tokens are added to $\tokens_i$, and thus  $|\tokens_i|$ is strictly
increasing. Because $|to\_color_i|$ remains unchanged, the condition
$|\tokens_i|<|to\_color_i|$ necessarily  becomes false at some point, 
which proves that the loop terminates.\\

Just after the loop,  the invariant is still true. In particular Assertion
(\ref{eq:colormapparentnonempty}) and  Assertion (\ref{eq:colorinonempty}) 
show that both the sets $\colors_i$ and $color\_map_i[parent_i]$ are not
empty when $p_i$ exits the \while loop.

Finally, due to to the fact that the message \ccolor$_j$() is well-formed, 
it  follows from M3 that we have 
$\colors_i \cap color\_map_i[parent_i] =\emptyset$ after line~\ref{MC-07}. 
As colors are added neither to  $\colors_i$, nor to   $color\_map_i[parent_i]$ 
in the loop, their intersection remains empty, 
which concludes  the  proof of the lemma.
\renewcommand{\toto}{lemma:while-loop-termination}
\end{proofL} 

\begin{lemma}
\label{lemma:well-formed-message}
All messages \ccolor$()$ broadcast at line~{\em\ref{MC-27}} are well-formed. 
\end{lemma}

\begin{proofL} 
To broadcast a message \ccolor$()$, a process $p_i$ must be in  
local state $1$ (line~\ref{MC-27}). This means that $p_i$
executed line~\ref{MC-21}, and consequently  previously received a
message \ccolor$(sender,cl\_map,max\_cl)$ 
that caused $p_i$ to execute  lines~{\ref{MC-05}-\ref{MC-23}}.\\

Let us first assume that \ccolor$()$ is well-formed. 
It then follows from~ Lemma~\ref{lemma:while-loop-termination} that $p_i$ 
exits the while loop,  and each of $\colors_i$ and $color\_map_i[parent_i]$ 
is not empty (A), and they have an empty intersection (B).
When considering the message  \ccolor$(id_i,color\_map_i,max\_cl_i)$
broadcast by $p_i$ we have the following.
\begin{itemize}
\vspace{-0.2cm}
\item M1 follows from the fact that the entries of the 
dictionary data structure created by $p_i$ are: 
$color\_map_i[parent_i]$ (line~\ref{MC-05}),
$color\_map_i[id_i]$ (line~\ref{MC-07}),
and $color\_map_i[id]$ for each 
$id\in to\_colors_i= \neighbors_i\setminus\{parent_i\}$
(lines~\ref{MC-06} and~\ref{MC-20}), and the observation 
that no entry is ever removed from $color\_map_i$ is the rest of the code.
\vspace{-0.2cm}
\item M2 follows from (A) for $color\_map_i[parent_i]$ and
  $color\_map_i[id_i]$, from line~\ref{MC-19} for the identities in
  $to\_colors_i=\neighbors_i\setminus\{parent_i\}$ (due to
  $|tokens_i|\geq |to\_colors_i|$ when line~\ref{MC-19} is executed,
  and the non-intersection requirement of the $tk[id]$ sets, no
  $tk[id]$ is empty), and from the observation that
    $color\_map_i$ is not modified between the end of line~\ref{MC-20}
    and the broadcast of line~\ref{MC-27}. This last claim is derived
    from the fact that $color\_map_i$ is only modified when messages
    are received, and that neither $p_i$'s parent nor $p_i$'s children
    are in states that allow them to send messages while $p_i$ is
    transitioning from line~\ref{MC-20} to line~\ref{MC-27}.
\vspace{-0.2cm}
\item Similarly M3 follows
\begin{itemize}
\vspace{-0.2cm}
\item for  $id=parent_i$: 
from (B) and the fact that  $color\_map_i[parent_i]$ never increases,
\vspace{-0.1cm}
\item  for $id\in to\_color_i=\neighbors_i\setminus\{parent_i\}$:
from the fact that, due to lines~\ref{MC-10} and~\ref{MC-12},  
at line~\ref{MC-19} $\tokens_i$ contains no token whose color belongs 
to  $\colors_i$, from which we have $tk[id]\cap color\_map_i[id_i]=\emptyset$ 
for any $id\in to\_color_i$.  
\end{itemize} 
\vspace{-0.2cm}
\item M4 follows from the construction of $\tokens_i$.
This construction ensures that, for any color $c$,   
$\tokens_i$ contains at most $m$ tokens with color $c$   
(line~\ref{MC-10},~\ref{MC-13}, and~\ref{MC-15}). 
\vspace{-0.2cm}
\item M5 is an immediate consequence of the assignment  
$max\_cl_i\leftarrow \max(max\_cl,\sigma_i)$ at line~\ref{MC-15}. 
\vspace{-0.2cm}
\item M6 follows from the following observations:
\begin{itemize}
\vspace{-0.1cm}
\item for $id\in\{id_i,parent_i\}$: from 
$max\_cl \leq max\_cl_i$ (line~\ref{MC-08}) and the fact that  
the message \ccolor$()$ received by $p_i$ is well-formed (hence 
$color\_map_i[id_i] \cup color\_map_i[parent_i] \subseteq [0..(max\_cl-1)]),$
\vspace{-0.1cm}
\item  for $id\in to\_color_i=\neighbors_i\setminus\{parent_i\}$:
from the fact that $\tokens_i$ contains only tokens whose color is in 
$[0..(max\_cl_i-1)]$ (line~\ref{MC-10}). 
\end{itemize} 
\end{itemize}

The previous reasoning showed that, if a process receives a 
well-formed message \ccolor$(),$ executes lines~\ref{MC-05}-\ref{MC-23}
and line~\ref{MC-27},  the message  
\ccolor$(id_i,color\_map_i,max\_cl_i)$ it will broadcast at this line is 
well-formed.  Hence, to show that all messages  broadcast at line~\ref{MC-27}
are well-formed, it only remains to show that the message 
\ccolor$(id_r,color\_map_r,max\_cl_r)$  broadcast by the root $p_r$ is 
well-formed. Let us remember that $\neighbors_r$ is a constant defined 
by the structure of the tree, and $parent_r=id_r\notin \neighbors_r$.

Let us notice that the message \ccolor$(id,cl\_map,max\_cl)$,
that $p_r$ sends to itself at line~\ref{MC-02},is not well-formed. 
This is because,  $cl\_map[id]$ is not defined for  $id\in \neighbors_i$.
When $p_r$ receives this message we have the following after 
line~\ref{MC-10}:\\
\centerline{
$|\tokens_r|= m\times |\sigma_r| -m= m \times (|\sigma_r| -1)
                   = m\lceil\frac{\Delta_r}{m}\rceil\geq \Delta_r,$}
from which we conclude 
$|\tokens_r|\geq \Delta_r = |to\_colors_r| =|neighbors_i|$.
Hence, $p_r$ does not execute the loop body, and proceeds to 
lines~\ref{MC-19}-\ref{MC-20} where it defines the entries 
$color\_map_r[id]$ for $id\in to\_colors_r=neighbors_r$. 
A  reasoning similar to the previous one shows that the message
\ccolor$(id_r,color\_map_r,max\_cl_r)$ broadcast by $p_r$ 
at line~\ref{MC-27} satisfies 
the properties M1-M6, and is consequently well-formed. 
(The difference with the previous reasoning lies 
in the definition of the set   $to\_colors_i$ 
which is equal to $\neighbors_i\setminus \{parent_i\}$ for $p_i\neq p_r$, 
and  equal to $\neighbors_r$ for $p_r$.) 
\renewcommand{\toto}{lemma:well-formed-message}
\end{proofL} 

\begin{lemma}
\label{lemma:no-empty-color-set}
If a process $p_i$ computes a color set ($\colors_i$), this set is not empty.
\end{lemma}

\begin{proofL}
Let us first observe that, if a process $p_i\neq p_r$ 
receives a message \ccolor$(-,cl\_map,-)$, 
the previous lemma means that this message is well-formed, 
and  due to property M2,  its field $cl\_map[id_i]$ is not empty, from 
which follows that the initial assignment of a value to  
$color\_map_i[id_i]\equiv \colors_i$ is a non-empty  set. 
Let us also observe, that, even if it is not well-formed the message 
\ccolor$(-,cl\_map,-)$ received by the root satisfies this property. 
Hence, any process that receives a message \ccolor$()$
assigns first a non-empty value to $color\_map_i[id_i]\equiv \colors_i$.

Subsequently, a color can only be suppressed from  
$color\_map_i[id_i]\equiv \colors_i$ at line~\ref{MC-24} 
when $p_i$ receives a message \ccolor$()$ from one of its children. 
If $p_i$ is a leaf, it has no children, and consequently 
never executes line~\ref{MC-24}. So, let us assume that $p_i$ is not a leaf
and receives a message \ccolor$(id_j,cl\_map,-)$ from one of its 
children $p_j$. In this case $p_i$  previously broadcast at line~\ref{MC-27} 
a message \ccolor$(id_i,color\_map_i,-)$ 
that was received by $p_j$ and this message is well-formed 
(Lemma~\ref{lemma:well-formed-message}). 

A  color $c$ that is suppressed at line~\ref{MC-24} when $p_i$ processes 
\ccolor$(id_j,cl\_map,-)$ is such that
$c\in \colors_i$ and $c\notin cl\_map[id_i]$. 
$cl\_map[id_i]$ can be traced back to the local variable
$color\_map_j[id_i]$ used by $p_j$ to broadcast \ccolor$()$ at 
line~\ref{MC-27}. Tracing the control flow further back,
$color\_map_j[id_i]$ was initialized by $p_j$ to $color\_map_i[id_i]$
(line~\ref{MC-05}) when $p_j$  received the well-formed
message \ccolor$()$ from $p_i$.When processing \ccolor$()$ received 
from $p_i$, process $p_j$ can  suppress colors from 
$color\_map_j[id_i]$ only at line~\ref{MC-16}, where it
suppresses colors starting from the greatest remaining color.
We have the following.
\begin{itemize}
\vspace{-0.2cm}
\item If $p_i$ is not the root, the message \ccolor$()$ 
it received was well-formed (Lemma~\ref{lemma:well-formed-message}). 
In this case, it follows from the  proof of 
Lemma~\ref{lemma:while-loop-termination} 
that it always remains at least one color in  $color\_map_j[id_i]$. 
\vspace{-0.2cm}
\item If $p_i=p_r$, its set $\colors_r$ is a singleton 
(it ``received'' \ccolor$(id_r, cl\_map_r,-)$ 
where $cl\_map_r$ has a single entry, namely  $cl\_map_r[id_r]=\{1\}$).
When $p_j$ computes $\tokens_j$ (line~\ref{MC-10}) we have \\
$|\tokens_j| =  m \times \max(|\sigma_r|,|\sigma_j|) -m
             = m\lceil\frac{\max(\Delta_r,\Delta_j)}{m}\rceil
            \geq   \max(\Delta_r,\Delta_j)\geq \Delta_j=|to\_colors_j|,$\\
from which follows that $|\tokens_j|\geq |to\_colors_j|=|neighbors_j|-1$.
Hence, $p_j$ does not execute the loop, and consequently 
does not modify $color\_map_j[id_r]$. 
\end{itemize}
Consequently, the smallest color of 
$colors_i\equiv color\_map_i[id_i]$  is never 
withdrawn from $color\_map_j[id_i]$. It follows that, at line~\ref{MC-24},  
$p_i$ never withdraws its smallest color from the set $color\_map_i[id_i]$. 
\renewcommand{\toto}{lemma:no-empty-color-set}
\end{proofL} 

\begin{lemma}
\label{lemma:Noconflict-multi-Coloring}
If $p_i$ and $p_j$ are neighbors $\colors_i \cap \colors_j= \emptyset$.
\end{lemma}

\begin{proofL}
As all color sets are initialized to $\emptyset$, the property is initially 
true. We show that, if a process receives a message \ccolor$()$, 
the property remains  true. As \tterm$()$ messages do not modify 
the coloring---lines~\ref{MC-30}-\ref{MC-35}---they do not need to be 
considered.

Let us consider two neighbor processes $p_i$ and $p_j$, which computes 
their color sets  (if none or only one of $p_i$ and $p_j$
computes its color set, the  lemma is trivially satisfied). 
As the network is a tree, one of them  is the parent of the other. 
Let $p_i$ be the parent of $p_j$. 

Process $p_i$ broadcast a message \ccolor$(-,cl\_map,-)$ at line~\ref{MC-27} 
in which the set  $cl\_map[id_j]$ is  $color\_map_i[id_j]$, as computed at 
line~\ref{MC-19}. If this message is received by $p_j$, this set will in 
turn be assigned to $color\_map_i[id_j]$ at $p_j$.
As this message is well-formed (Lemma~\ref{lemma:well-formed-message}),  
we therefore have $color\_map_i[id_i]\cap  color\_map_j[id_j]=\emptyset$
(Property M3 of a well-formed message).  
Then, while $p_i$ can be directed to suppress colors from $color\_map_i[id_i]$
at line~\ref{MC-24}, it never adds a color to this set. 
The same is true for $p_j$ and $color\_map_j[id_j]$. It follows that 
the predicate $color\_map_i[id_i]\cap  color\_map_j[id_j]=\emptyset$
can never be invalidated.
\renewcommand{\toto}{lemma:Noconflict-multi-Coloring}
\end{proofL}

\begin{lemma}
\label{lemma:multi-Nocollision-Coloring}
$\forall i, \forall c:
 |\{ j: j \in \neighbors_i \wedge c\in \mathit{\colors_j} \} | \leq m$. 
\end{lemma} 

\begin{proofL}
The property is initially true.  
We show that it remains true when processes receive messages.  

Let us consider a process $p_i$ that broadcasts a message \ccolor$()$. 
Due to the fact that such  messages are  broadcast only at line~\ref{MC-27}, 
it follows from Lemma~\ref{lemma:well-formed-message} that 
the   message \ccolor$(id_i,cl\_map,-)$ broadcast by $p_i$ is  well-formed. 
Hence it satisfies property M4. 
When processing this message
\begin{itemize}
\vspace{-0.2cm}
\item[A] each child $p_j$  of $p_i$ adopts $cl\_map[id_j]$ as its 
initial color set  and assigns it to $color\_map_j[id_j]$;
\vspace{-0.2cm}
\item[B] $p_i$'s parent $p_k$ uses $cl\_map[id_k]$ to 
update $color\_map_k[id_k]$ at line~\ref{MC-24} such that 
$color\_map_k[id_k]$ $\subseteq$ $cl\_map[id_k]$.
\vspace{-0.2cm}
\end{itemize}
(A), (B), and M4 imply that just after $p_i$'s neighbors have processed 
$p_i$'s message, the lemma holds. As already seen in the proof of 
other lemmas, $color\_map_j[id_j]$ may subsequently decrease, but never 
increases: 
colors can be suppressed from $color\_map_j[id_j]$ (line~\ref{MC-24})
but never added to it. And the same is true at $p_i$ for its set of 
colors $color\_map_i[id_i]$, and at its parent $p_k$ for $color\_map_k[id_k]$. 
It then follows that 
$|\{ j: j \in \neighbors_i \wedge c\in \mathit{\colors_j} \} | \leq m$ 
throughout the execution of the algorithm, 
which concludes the proof of the lemma. 
\renewcommand{\toto}{lemma:multi-Nocollision-Coloring}
\end{proofL}

\begin{lemma}
\label{lemma:collision-freedom}
Algorithm~{\em\ref{fig:tree-multicoloring-algorithm}} is collision-free. 
\end{lemma}

\begin{proofL}
We have to show that no process can have more than $m$ of its
neighbors that broadcast during the same round. Initially, all
processes are in state $0$. Let us consider a process $p_i$
and assume that one of its neighbors $p_j$ is broadcasting a
message. Let us further assume that this message is of type \ccolor().
\begin{itemize}
\vspace{-0.2cm}
\item If $p_j$ is $p_i$'s parent, $p_j$'s \ccolor() message is the first 
message received by $p_i$, and both $p_i$ and its children ($p_i$'s 
remaining neighbors) are in state $0$, and hence silent. 
There is no collision at $p_i$.
\vspace{-0.2cm}
\item 
If $p_j$ is one of $p_i$'s children, the value $slot\_span_j$ used by
$p_j$ at line~\ref{MC-26} is equal to $max\_cl$ contained in the
message \ccolor$(-,-,max\_cl)$ first received by $p_j$ from $p_i$.
Because of Lemma~\ref{lemma:well-formed-message}, this message is
well-formed, and consequently satisfies property M6.  Any other child
$p_\ell$ of $p_i$ broadcasting during this round will have received the
same first message, and will therefore be using the same 
$slot\_span_\ell = max\_cl$ value. 
It follows from 
Property M6, the assignment of line~\ref{MC-07} 
executed by any child $p_\ell$ (of $p_i$)  that received the message, 
and the fact that its set $\colors_\ell$ can only decrease after 
being first  assigned,  that
$\colors_\ell \subseteq [0 .. slot\_span_\ell-1 ] 
  \text{ for any child } p_\ell \text{ of } p_i$ (C).

Lemma~\ref{lemma:multi-Nocollision-Coloring}, Property (C), 
and the $\CLOCK$-based predicate defining
the rounds at which a process is allowed to broadcast
(line~\ref{MC-26}), imply that at most $m$ children of $p_i$ can
broadcast during the same round. If $p_i$ has a parent $p_k$
(i.e. $p_i$ is not the root), both $p_i$ and $p_k$ are in state $2$,
and hence $p_k$ is silent, proving the lemma. If $p_i$ is the root,
all its neighbors are its children, and the lemma also holds.
\end{itemize}
The same reasoning applies to the messages {\sc term}$()$ 
broadcast  by the  children of  $p_i$ and its parent.
\renewcommand{\toto}{lemma:collision-freedom}

\end{proofL} 


\begin{lemma}
\label{lemma:multi-Coloring}
Each process computes  a set of colors, and the root process 
knows when  coloring is terminated. 
\end{lemma}

\begin{proofL} 
Let us first observe that, due to Lemmas~\ref{lemma:while-loop-termination} 
and~\ref{lemma:well-formed-message},
no  process $p_i\neq p_r$  can loop forever inside the {\bf while} loop 
(lines~\ref{MC-11}-\ref{MC-18}), when  it receives its first message 
\ccolor$()$. 
The same was proved for the root $p_r$ at the end of the proof of 
Lemma~\ref{lemma:no-empty-color-set}.
Moreover, a process cannot block at line~\ref{MC-24} when it receives 
other messages  {\sc color}$()$  (one from each of its children). 
Hence,  no reception of a message \ccolor$()$ can prevent 
processes from terminating the processing of the message.  
The same is trivially true for the  processing of a message {\sc term}$()$. \\ 

Let us first show that each process obtains a non-empty set of colors.
To this end, we show that each non-leaf process broadcasts a message
\ccolor$()$. 
\begin{itemize}
\vspace{-0.2cm}
\item 
When the root process $p_r$ receives the external message {\sc
  start}$()$, it ``simulates the sending to itself'' of the message
\ccolor$(id_r,color\_map_r,\sigma_r)$, where the dictionary data
structure $color\_map_r$ has a single element, namely,
$color\_map_r[id_r]=\{1\}$.  The root $p_r$ executes consequently the
lines~\ref{MC-25}-\ref{MC-23}, during which it obtains a color
($color\_map_r[id_r]=\{1\}$, line~\ref{MC-07}), and computes a set of
proposed colors $color\_map_r[id_j]$ for each of its children $p_j$
(lines~\ref{MC-19}-\ref{MC-20}). It then progresses to the local
non-waiting state $1$ (line~\ref{MC-21}).  Hence, during the first
round, it broadcasts to its neighbors the message
\ccolor$(id_r,color\_map_r,\sigma_r)$.Because the algorithm
  is conflict- and collision-free (Lemmas~\ref{lemma:noconflict}
  and~\ref{lemma:Noconflict-multi-Coloring}), this message is received
  by all the root's neighbors.
\vspace{-0.2cm}
\item Let us now consider a process $p_i$ that receives a message
  \ccolor$(sender,color\_map,max\_cl)$ for the first time.  It follows
  from Lemma~\ref{lemma:no-empty-color-set} that $p_i$ starts
  computing a non-empty set $\colors_i$ and enters the waiting state
  $1$ (line~\ref{MC-21}).  Finally, 
as  $\colors_i$ $\subseteq$ $[0  .. slot\_span_i $ $-1]$, and 
 $\CLOCK$ never stops increasing,  the predicate of line~\ref{MC-26}
is eventually satisfied. It follows
$p_i$ broadcasts the message\ccolor$(id_i,color\_map_i,max\_cl_i)$. 
As above all of $p_i$'s  neighbors will receive this message.

It follows that \ccolor$()$ messages flood the tree from the root to
the leaves. 

Moreover, when a process $p_i$ has received a message  \ccolor$()$
from each of its neighbors (children and parent),  it 
has obtained the final value of its color set $color\_map_i[id_i]=\colors_i$. 
Due to lemma~\ref{lemma:no-empty-color-set}, this set is not empty, 
which concludes the first part of the proof.  
\end{itemize}

Let us now show that the root learns  coloring termination. 
This relies on the messages \tterm$()$. 
As previously,  due to Lemma~\ref{lemma:conflict-freedom}
and  Lemma~\ref{lemma:collision-freedom}, these messages entail neither 
message conflicts nor message collisions. 

Let us observe that each leaf process enters the non-waiting state $3$.
When the predicate of line~\ref{MC-26} is satisfied at a leaf $p_\ell$
(this inevitably occurs),  
this process broadcasts the message \tterm$()$ to its parent $p_i$.
Then, when $p_i$ has received a message  \tterm$()$ from
each of its children, it broadcasts \tterm$()$ to its own parent.
This sequence repeats itself on each path from a leaf to the root. 
When the root has received a message  \tterm$()$
from each of its children, it learns termination (line~\ref{MC-34}), 
which concludes the proof of the lemma. 
 \renewcommand{\toto}{lemma:multi-Coloring}
\end{proofL}

\begin{lemma}
\label{lemma:multi-K-bound}
$|\bigcup_{1\leq i\leq n} \mathit{\colors_i}| = \lceil\frac{\Delta}{m}\rceil+1$.
\end{lemma}

\begin{proofL}
Let $p_r,p_a, \cdots, p_\ell$ be a  path in the tree
starting at the root $p_r$ and ending at a leaf  $p_\ell$. 
It follows from 
\begin{itemize}
\vspace{-0.2cm}
\item 
the content of the parameter $max\_cl$ of
the messages \ccolor$(sender,cl\_map, max\_cl)$ broadcast 
along this path of the tree (broadcast
at line~\ref{MC-27} and received at line~\ref{MC-03}), and
\vspace{-0.2cm}
\item 
the assignment of ${\sf max}(max\_cl,\sigma_i)$ to $max\_cl_i$
at line~\ref{MC-08},
\end{itemize}
\vspace{-0.2cm}
that $max\_cl_\ell = {\sf max}(\sigma_r,\sigma_a,\cdots, \sigma_\ell)$.
Let $p_{\ell 1}$, ..., $p_{\ell x}$ be the set of leaves of the tree.
It follows that  ${\sf max}(max\_cl_{\ell 1}, \cdots, max\_cl_{\ell x})$
=  ${\sf max}(\sigma_1, \cdots, \sigma_n)$, i.e., the value $max\_cl$ 
carried by any message is $\leq  \lceil\frac{\Delta}{m}\rceil+1$.

The fact that a process $p_i$ uses only colors in $[0..(max\_cl_i-1)]$, 
combined  with  Theorem~\ref{theo:lower-bound}  implies the lemma.
The algorithm is consequently optimal with respect to the number of colors.  
\renewcommand{\toto}{lemma:multi-K-bound}
\end{proofL}

\begin{theorem}
\label{theo:algo-2} 
Let $K=\lceil\frac{\Delta}{m}\rceil+1$.
Algorithm {\em{\ref{fig:tree-multicoloring-algorithm}}} is a 
{\em C2$m$-free} algorithm, which solves  
{\em CCMC($n,m,K,\geq 1$)} in tree networks.
Moreover, it is optimal with respect to the value of $K$. 
\end{theorem}

\begin{proofT}
The proof that Algorithm 1 is C2$m$-free follows from 
Lemma~\ref{lemma:noconflict} and Lemma~\ref{lemma:collision-freedom}. 
The proof that it satisfies the Conflict-freedom, Collision-freedom, and 
Efficiency properties defining  the CCMC($n,m,K,\geq 1$) problem
follows from Lemmas
\ref{lemma:while-loop-termination}-\ref{lemma:multi-Nocollision-Coloring}, 
and Lemma~\ref{lemma:multi-Coloring}.
The proof of its optimality with respect to $K$ follows from  
Lemma~\ref{lemma:multi-K-bound}.
\renewcommand{\toto}{theo:algo-2}
\end{proofT}

\section{Conclusion}
\label{sec:conclusion}
The paper first introduced a new vertex coloring problem (called
CCMC), in which a process may be assigned several colors in such a way
that no two neighbors share colors, and for any color $c$, at most
$m$ neighbors of any vertex share the color $c$.  This coloring
problem is particularly suited to assign rounds (slots) to processes
(nodes) in broadcast/receive synchronous communication systems with
communication or local memory constraints.  Then, the paper presented
a distributed algorithm which solve this vertex coloring problem for
tree networks in a round-based programming model with
  conflicts and (multi-frequency) collisions. This algorithm is
optimal with respect to the total number of colors that can be used,
namely it uses only $K=\lceil\frac{\Delta}{m}\rceil+1$ different
colors, where $\Delta$ is the maximal degree of the graph.

It is possible to easily modify the coloring problem CCMC to
express constraints capturing specific
broadcast/receive communication systems. As an example,
suppressing the conflict-freedom constraint  and  
weakening the collision-freedom constraint into
\begin{equation}
  \forall i, \forall c: |\{ j: (id_j \in \mathit{neighbors_i}\cup
  \{id_i\}) \wedge (c\in \colors_j) \}| \leq m,
  \label{eq:ballonpi}
\end{equation}
captures bi-directional communication structures encountered in some
practical systems in which nodes may send and receive on distinct
channels during the same round. Interestingly, solving the coloring problem
captured by~(\ref{eq:ballonpi}) is equivalent to solving distance-2
coloring in the sense that a purely local procedure (i.e., a procedure
involving no communication between nodes) executed on each node can
transform a classical distance-2 coloring into a multi-coloring
satisfying ~(\ref{eq:ballonpi}).
 More precisely, assuming a coloring $\col : V \mapsto [0 .. (K*m) -1]$ 
providing a distance-2 coloring with $K*m$ colors on a graph $G = (V,E)$, 
it is easy to show that the coloring (with one color per vertex) 
\begin{equation}
  \begin{array}{rrcl}
    \col' : & V & \mapsto     & [0\; ..\; K -1] \\
                    & x & \rightarrow & \col(x) {\sf ~mod~} K,
  \end{array}
\end{equation}
fulfills ({\ref{eq:ballonpi}) on $G$~(\footnote{This is because (a)
distance-2 coloring ensures that any vertex 
and its neighbors have different colors, and  (b) 
there are at most $m$  colors  $c_1,...,c_x \in  [0.. (K*m)-1]$
(hence $x\leq m$),  such that 
$(c_1  {\sf ~mod~} K) = \cdots= (c_x  {\sf ~mod~} K) =c \in [0..(K-1)]$.}).
 Since the distance-2 problem
  with $K*m$ colors is captured by CCMC$(n,1,K*m,1)$ (as discussed in
  Section~\ref{sec:problem}),  the proposed algorithm can also solve 
  the coloring condition captured by (\ref{eq:ballonpi}) on trees in 
  our computing model.

Moreover, from an algorithmic point of view, the proposed algorithm is
versatile, making it an attractive starting point to address other
related problems. For instance, in an heterogeneous network,
lines~\ref{MC-19}-\ref{MC-20} could be modified to take into account
additional constraints arising from the capacities of individual
nodes, such as their ability to use only certain frequencies.

Last but not least, a major challenge for future work consists in
solving the CCMC problem in general graphs. The new difficulty is then
to take into account cycles.


{}

\appendix

\section{Proof of Theorem~\ref{theo:multicoloring-condition}}
\label{proof:theo-multicoloring}

\begin{lemma}
\label{lemma:integer-arithmetic}
$K' > \left\lfloor\frac{\Delta_j}{m}\right\rfloor 
\Longleftrightarrow K'\times m > \Delta_j$.
\end{lemma}

\begin{proofL}
 From $K' > \lfloor\frac{\Delta_j}{m}\rfloor$ we can derive the following. 
Because  $K'$ is an integer, we have 
$K'\geq \lfloor\frac{\Delta_j}{m}\rfloor + 1$. 
Because $\lfloor x \rfloor > x -1$, we obtain 
$K' >\frac{\Delta_j}{m} -1 + 1$, from which we conclude 
$K' \times m  >\Delta_j.$
Conversely, if we have $K'\times m > \Delta_j$, then
$K' >\frac{\Delta_j}{m}$, and, because $x \geq \lfloor x \rfloor$, 
we obtain $ K' >\frac{\Delta_j}{m}  \geq
\lfloor\frac{\Delta_j}{m}\rfloor$,  which concludes the proof of the lemma.
\renewcommand{\toto}{lemma:integer-arithmetic}
\end{proofL}

\begin{theorem-repeat}{theo:multicoloring-condition}
Let $K= \lceil \frac{\Delta}{m} \rceil +1$. 
CCMC$(n,m,K, >1)$
can be solved on a tree of maximal degree $\Delta$, if and only if
\vspace{-0.2cm}
$$\exists i :  K > \max
\left(\; \left\{ \left\lceil \textstyle\frac{\Delta_i}{m}\right\rceil \right\} 
\cup \left\{ \left\lfloor \left. \textstyle\frac{\Delta_j}{m}\right\rfloor \;
\right|\; p_j \in \neighbors_i \right\}\;\right) + 1.$$
\end{theorem-repeat}

\begin{proofT} 
The terms ``process'' and ``vertex'' are considered here as synonyms. 
To simplify notation, we consider in the following that $id_i=i$. 
Let us first notice that if follows from its definition that $K\geq 2$. \\

\noindent
{\sl Proof of the ``if direction''.} \\
The proof of this direction consists in a sequential algorithm that associates 
two colors to a process $p_i$ whose position in the tree satisfies the 
previous predicate.  

Algorithm~\ref{fig:tree-DF-algorithm} is a sequential algorithm
solving CCMC$(n,m,\lceil \frac{\Delta}{m} \rceil +1,1)$. Using the
control flow defined by a simple depth-first tree traversal algorithm,
it takes two input parameters, a process $p_j$, and its color.  Then,
assuming a coloring of both $p_j$ and its parent, it recursively
colors the vertices of the tree rooted at $p_j$.  The initial call is
${\sf DF\_MColoring}(i,0)$ where $p_i$ is a vertex satisfying the
predicate stated in the theorem, and $0$ the color assigned to it.
The function ${\sf parent\_color}(j)$ returns the color of the parent
of $p_j$ if $p_j\neq p_i$ and returns no value if $p_j=p_i$. 
Let us notice that, except for $p_j=p_i$, ${\sf parent\_color}(j)$ is 
called only after the parent of $p_j$ obtained a color.

\begin{algorithm}[h!]
\centering{
\fbox{
\begin{minipage}[t]{150mm}
\footnotesize
\renewcommand{\baselinestretch}{2.5}
\resetline
\begin{tabbing}
aaaA\=A\=aA\=aaA\=aaA\=aaA\kill

{\bf  procedure} ${\sf DF\_MColoring}(j,c)$ {\bf  is} \\

\line{MC-01}  \> \> 
$color_i\leftarrow c$;\\
\line{MC-02}  \> \> 

{\bf if} $|\neighbors_j|>1$ {\bf then}\\

\line{MC-03}  \> \>  \> 
$\tokens \leftarrow$ \=
$\{$ $m$ colored tokens for each 
color in $\{0,1,\cdots,(K-1) \}\setminus 
            \{color_j, {\sf parent\_color}(j)\}$ $\}$\\

 \>\>\>  \> 
$\cup$ $\{$ $(m-1)$ tokens with color ${\sf parent\_color}(j)\}$;\\

\line{MC-04}  \> \>  \> 
{\bf for} \= {\bf each} 
        $p_k\in (\neighbors_j\setminus \{parent_j\})$ {\bf do}
       $~~$ \%this loop is executed $(\Delta_j-1)$ times \% \\

\line{MC-05}  \> \>  \> \> 
$token\leftarrow$ a smallest token in $\tokens$; 
suppress $token$ from $\tokens$;\\

\line{MC-06}  \> \>  \>  \>  
 ${\sf DF\_MColoring}(k,\mbox{ color of } token)$ \\

\line{MC-07}  \> \>  \> {\bf end for}\\

\line{MC-08}  \> \>  {\bf end if}.

\end{tabbing}
\normalsize
\end{minipage}
}
\caption{Sequential multi-coloring of a tree with a depth-first  traversal 
algorithm (code for $p_i$)}
\label{fig:tree-DF-algorithm}
}
\end{algorithm}

A call to  ${\sf DF\_MColoring}(j,c)$ works as follows. 
First, color $c$ is assigned to $p_j$. 
If $p_j$ has a single neighbor (its parent, which issued this call),
the current procedure call terminates. Otherwise, 
the current invocation computes  the multiset of colored tokens which
includes (a) $m$ identical tokens for each possible color, 
except the colors of $p_j$ and its parent, and  (b) $(m-1)$ identical tokens
with the color of $parent_j$ (line~\ref{MC-03}). 
Let us notice that all the colored tokens are ordered by their color number, 
hence the notion of a ``token with a smallest color'' is well-defined. 
Then, for each of $p_j$'s neighbor $p_k$ (except its parent),
taken one after the other (line~\ref{MC-04}),  a token with a smallest 
color is selected for $p_k$ (which will inherit the corresponding color)
and withdrawn from the multiset $\tokens$ (hence, this token  can
no longer  be used to associate a color to another neighbor of  $p_j$, 
line~\ref{MC-04}). 
Due to line~\ref{MC-03}, 
the multiset $\tokens$ used to assign colors to $p_j$'s neighbors, is such
that  $|\tokens|= (K-2)\times m + (m-1)=  (K-1)\times m -1$ (Property P1).
(The factor $(K-2)$ comes from the fact the colors are in the color set 
$\{0,1,\cdots,(K-1)\}\setminus\{ color_j,{\sf parent\_color}(j)\}$. 
The term $(m-1)$ comes from the fact that $color_i=0$ is already used once for
the neighbor  $p_i$.) 
It follows that  the loop is well-defined. The subtree rooted at $p_k$ is then
depth-first recursively  colored (line~\ref{MC-05}). 

It follows that (a) no two neighbors can be assigned the same color 
(line~\ref{MC-03}), 
(b) each process is assigned a color as small as 
possible (line~\ref{MC-04}), and (c) at most $m$ neighbors of a process 
can be  assigned  the same color (lines~\ref{MC-02} and~\ref{MC-05}). \\

We now show that, given the previous coloring, it is possible to
assign  (at least) one more color to (at least) the root $p_i$. 
The set of $K$ colors used in Algorithm~\ref{fig:tree-DF-algorithm} is
the set $\{0,1,\cdots,(K-1)\}$.  Let us consider any vertex $p_j$,
which is a neighbor of $p_i$.  Due to the property stated in the
theorem, we have $K> \lfloor \frac{\Delta_j}{m}\rfloor +1$, which, due
to Lemma~\ref{lemma:integer-arithmetic} translates as 
$(K-1)\times m > \Delta_j$ (P2).

As $color_i=0$ and $ K\geq 2$, we  have $color_i\neq K-1$.
 Moreover, we also have $color_j\neq K-1$. 
This follows from the assumption  $K >\lceil\frac{\Delta_i}{m}\rceil+1$, 
and the fact that, to color $p_i$ and its neighbors,
Algorithm~\ref{fig:tree-DF-algorithm} uses 
$\lceil\frac{\Delta_i}{m}\rceil+1\leq K-1$
colors, namely the  color set $\{0,\cdots,(K-2)\}$.


When executing  ${\sf DF\_MColoring}(j,c)$ for a neighbor $p_j$ of $p_i$,
Algorithm~\ref{fig:tree-DF-algorithm} executes  $(\Delta_j -1)$ times 
the body of the ``for'' loop  (lines~\ref{MC-04}-\ref{MC-06}),  
(once for each neighbor of $p_j$, except $p_i$, which has already been 
assigned  a color). 
It follows from  (P2) that $(K-1)\times m -1 > \Delta_j -1$. 
Combined with (P1) we obtain $|tokens|  > \Delta_j -1$, from which we 
conclude that   $tokens \neq \emptyset$  is always true. 
It is consequently a loop invariant in each call related to a neighbor 
$p_j$ of $p_i$. It follows that $tokens$ always contains 
a colored token with the highest color, namely $(K-1)$. 
This color can consequently be assigned to $p_i$, in addition of color $0$,
without violating the conflict-freedom, $m$-collision-freedom, 
and efficiency defining the CCMC problem solved by 
Algorithm~\ref{fig:tree-DF-algorithm}. This concludes the proof
of the  ``if'' part of the theorem.~\\

\noindent
{\sl Proof of the ``only if direction''}.  \\
  \newcommand{\multip}[1]{\textbf{1}_{#1}}
  \newcommand{\myset}{\mathit{set}}
In the following we  use the following notations, where $M$ is a multi-set.
\begin{itemize}
\vspace{-0.2cm}
\item $|M|$ is the size of $M$ (in the following all sets and multisets 
are finite),
\vspace{-0.2cm}
\item 
$\myset(M)$ is the underlying set of $M$, the set of elements present 
at least once in $M$,
\vspace{-0.2cm}
\item $\multip{M}(x)$ is the multiplicity of an element $x$ in $M$. 
By construction we have
\begin{align}
      \multip{M}(x) \geq 1 &\Longleftrightarrow x \in \myset(M) 
        ~~{\mbox{ and }}~~
      |M|= \sum_{x\in \myset(M)} \multip{M}(x).
\label{eq:cardimulti}
\end{align}
\end{itemize}

\noindent
If $A$ and $B$ are two multisets, 
$A \uplus B$ is the multiset union of $A$ and $B$. In particular we have:
   \label{eq:cardimultiunion}
\begin{align}
   |A \uplus B| =|A|+|B|  ~~{\mbox{ and }}~~
     \multip{A \uplus B}(x) = \multip{A}(x) + \multip{B}(x).
 \label{eq:cardimultiunion}
 \end{align}   
We consider a set $S$ as a special case of a multi-set in which all
elements of $S$ have a multiplicity of 1: $x \in S
\Longleftrightarrow \multip{S}(x) = 1$.\\

\noindent  
Let us assume that CCMC($n,m,K, >1$) can be solved on a tree, such that 
at least one process, i.e. $p_i$, is allocated more than one color:
  \begin{equation}\label{eq:morethan2oni}
    |\colors_i| \geq 2.
  \end{equation}
  \noindent 
\noindent
For ease of exposition, and without loss of generality, we assume  all 
other processes are allocated only one color:
  \begin{equation}\label{eq:cjonecolor}
    \forall j \neq i: |\colors_j| = 1.
\end{equation}
Let $C_{\neighbors_i}$ denote the multiset of colored tokens allocated to the 
neighbors of $p_i$:
  \begin{equation}\label{eq:defcni}
    C_{\neighbors_i} = \mathop{\biguplus}_{p_j\in \neighbors_i}\colors_j.
  \end{equation}
From (\ref{eq:cjonecolor}) and (\ref{eq:defcni}) we derive 
(by way of (\ref{eq:cardimultiunion}))
  \begin{equation}\label{eq:cni:equal:deltai}
    |C_{\neighbors_i}| = |\mathop{\biguplus}_{p_j\in \neighbors_i}\colors_j|=
      \sum_{p_j\in \neighbors_i} |\colors_j| = |\neighbors_i| = \Delta_i.
  \end{equation}%
This means that $\Delta_i$ colored tokens are needed to color the 
neighbors of $p_i$. 

\noindent
  Because the coloring solves CCMC($n,m,K, >1$), \emph{$m$-Collision-freedom}
 means that
  \begin{equation}\label{eq:mcollifreedom}
    \forall c \in \myset(C_{\neighbors_i}): \multip{C_{\neighbors_i}}(c) \leq m.
  \end{equation}
Using (\ref{eq:mcollifreedom}) in (\ref{eq:cardimulti}) applied 
to $C_{\neighbors_i}$ gives us
  \begin{align}
    |C_{\neighbors_i}| &= \sum_{c\in \myset(C_{\neighbors_i})} \multip{C_{\neighbors_i}}(c) 
    \leq |\myset(C_{\neighbors_i})| \times m,
  \end{align}
\noindent 
which yields, with (\ref{eq:cni:equal:deltai})
(required number of colors for $p_i$'s neighbors):
  \begin{align}
    |\myset(C_{\neighbors_i})| &\geq \frac{|C_{\neighbors_i}|}{m}
    \geq \frac{\Delta_i}{m}.\label{eq:geqdeltaioverm}
  \end{align}
Because $|\myset(C_{\neighbors_i})|$ is an integer, (\ref{eq:geqdeltaioverm}) 
implies that
  \begin{equation}\label{eq:cardcnigeqceil}
    |\myset(C_{\neighbors_i})| \geq \left\lceil \frac{\Delta_i}{m} \right\rceil.
  \end{equation}
Because the coloring solves  CCMC($n,m,K, >1$), it respects 
\emph{Conflict-freedom}, implying that
  \begin{align}
    \myset(C_{\neighbors_i}) \cap \colors_i &= \emptyset,
  \end{align}
\noindent 
and hence
  \begin{align}
    |\myset(C_{\neighbors_i}) \cup \colors_i| &= |\myset(C_{\neighbors_i})| + |\colors_i|,\\
    &\geq |\myset(C_{\neighbors_i})| + 2   &\text{using (\ref{eq:morethan2oni})},\\
    &\geq \left\lceil \frac{\Delta_i}{m} \right\rceil + 2                       
    > \left\lceil \frac{\Delta_i}{m} \right\rceil + 1. 
                &  \text{using  (\ref{eq:cardcnigeqceil})}
  \end{align}
  By definition $K \geq |\myset(C_{\neighbors_i}) \cup \colors_i|$, which yields
  \begin{equation}
    K > \left\lceil \frac{\Delta_i}{m} \right\rceil + 1 \label{eq:okforpi}
  \end{equation}
which concludes the first part of the proof on the ``only if'' direction.

\noindent
Let us now turn to the neighbors of $p_i$.
 For $p_j \in \neighbors_i$ we consider similarly to $p_i$ the set of colored 
tokens allocated to $p_j$'s neighbors (which include $p_i$):
  \begin{equation}\label{eq:defcnj}
    C_{\neighbors_j} = \mathop{\biguplus}_{p_k\in \neighbors_j}\colors_k.
  \end{equation}
Hence (as for $p_i$) we have:
  \begin{equation}\label{eq:cnj:equal:sum}
    |C_{\neighbors_j}| = \sum_{p_k\in \neighbors_j} |\colors_k|.
  \end{equation}

\noindent
Contrary to $p_i$ however, all of $p_j$'s neighbors do not have only one 
color allocated: $p_i$ has at least two,  by assumption. This yields
  \begin{align}
    |C_{\neighbors_j}| &= |\colors_i| + 
                        \sum_{p_k\in \neighbors_j\setminus\{p_i\}} |\colors_k|,\\
    &\geq 2 + |\neighbors_j-1| \times 1 
      & \text{using (\ref{eq:morethan2oni}) and (\ref{eq:cjonecolor})},\\
    &\geq 2 + \Delta_j - 1
     \geq \Delta_j + 1.\label{eq:cnj:geq:deltaplus}
  \end{align}
  As for $p_i$, \emph{$m$-Collision-freedom} means that
  \begin{align}\label{eq:cnj:leq:myset:m}
    |C_{\neighbors_j}| & \leq |\myset(C_{\neighbors_j})| \times m.
  \end{align}
  (\ref{eq:cnj:geq:deltaplus}) and (\ref{eq:cnj:leq:myset:m}) yield
  \begin{align}\label{eq:cnjgeqdeltaplusone}
   |\myset(C_{\neighbors_j})| \times m &\geq \Delta_j + 1.
  \end{align}
  As for $p_i$ we have (\emph{Conflict-freedom})
  \begin{align}\label{eq:cnj:cap:empty}
    \myset(C_{\neighbors_j}) \cap \colors_j &= \emptyset,
  \end{align}
  \noindent leading to
  \begin{align}
    K &\geq |\myset(C_{\neighbors_j}) \cup \colors_j| & \text{by definition},\\
    &\geq |\myset(C_{\neighbors_j})| + |\colors_j| 
                    & \text{because of (\ref{eq:cnj:cap:empty})},\\
    &\geq |\myset(C_{\neighbors_j})| + 1 & \text{because of (\ref{eq:cjonecolor})},\\
    K -1 &\geq |\myset(C_{\neighbors_j})|.\label{eq:kminusone}
  \end{align}
  Injecting (\ref{eq:kminusone}) into (\ref{eq:cnjgeqdeltaplusone}) gives us
  \begin{align}
    (K - 1) \times m &\geq \Delta_j + 1,\\
    (K - 1) \times m &> \Delta_j,\\
    K - 1 &> \frac{\Delta_j}{m} \geq
             \left\lfloor\frac{\Delta_j}{m}\right\rfloor
      & \text{because of Lemma~\ref{lemma:integer-arithmetic}},\\
    K &> \left\lfloor\frac{\Delta_j}{m}\right\rfloor +1.\label{eq:okforpjaswell}
  \end{align}
(\ref{eq:okforpjaswell}) concludes 
the proof of  necessary condition to solve CCMC($n,m,K, >1$).
\renewcommand{\toto}{theo:multicoloring-condition}
\end{proofT}

\section{Defining the slots of the upper layer programming level}
When the root process claims termination, the other processes
are in their local state 4, but cannot exploit the multi-coloring assignment.
As indicated in the paragraph ``Using the colors'' (just before 
Section~\ref{sec:lower-bound}), to do that, they need 
to know the value $K=\lceil\frac{\Delta}{m}\rceil+1$.

The knowledge of $K$ can be brought to the root process by the messages
{\sc term}$()$, and then disseminated from the root to all the processes. 
Algorithm~\ref{fig:tree-multicoloring-algorithm} is slightly modified and 
enriched with Algorithm~\ref{fig:tree-parallel-coloring-termination}
to allow all processes to know the value of $K$. Modified lines are postfixed 
by a ``prime'', and new lines are numbered Nxy.

\begin{algorithm}[ht]
\centering{
\fbox{
\begin{minipage}[t]{150mm}
\footnotesize
\renewcommand{\baselinestretch}{2.5}
\begin{tabbing}
aaaaA\=A\=aaA\=aaA\=aaA\=aaA\kill

(\ref{MC-28}')  \> \>  \>{\bf else} 
 ${\sf broadcast}$ {\sc term}$(parent_i,id_i,kprime)$;  
 $state_i\leftarrow 4$ \\~\\

(\ref{MC-30}')  \>
{\bf when}  {\sc term}$(dest,id,ak)$ {\bf is received} {\bf do}\\

(\ref{MC-31}) \>\> {\bf if} $(dest \neq id_i)$ 
   {\bf then} discard the message
      (do not execute lines~\ref{MC-31}-\ref{MC-35}) {\bf end if};\\

(\ref{MC-32}) \>\> $colored_i \leftarrow to\_colored_i \cup \{ id\}$;\\

(N1) \>\>  $ak_i \leftarrow  {\sf max}(ak_i,ak);$  \\

(\ref{MC-33})   \>\> {\bf if} \= ($colored_i= \neighbors_i$)  \\

(\ref{MC-34}')  \>\>\> {\bf then if}  ($parent_i=id_i$) 
 {\bf then}   $state_i\leftarrow 5$
 {\bf else}   $state_i\leftarrow 3$  {\bf end if} \\

(\ref{MC-35})  \>\> {\bf end if}.\\~\\


(N2) \> {\bf when}  
$\big((\CLOCK {\sf~ mod~} ak_i) \in  color_i) 
                                 \wedge (state_i =5)\big)$ {\bf do}\\

(N3)  \>\>
 {\bf if}  $(|\neighbors_i|\neq 1)$  {\bf then} 
 ${\sf broadcast}$ {\sc end}$(id_i,k_i)$ {\bf end if}; 
 $state_i\leftarrow 6$.\\~\\


(N4) \> {\bf when} {\sc end}$(sender,k)$ {\bf is received} {\bf do} \\

(N5)  \>\> 
  {\bf if}  $(sender=parent_i)\wedge (state_i=4)$ \\

(N6)  \>\>\> 
 {\bf then}  $k_i \leftarrow {\sf max}(k_i,k)$; $state_i\leftarrow 5$ \\

(N7) \> \> {\bf end if}. 
 
\end{tabbing}
\normalsize
\end{minipage}
}
\caption{Obtaining the value $K=\lceil\frac{\Delta}{m}\rceil+1$
and informing all processes}
\label{fig:tree-parallel-coloring-termination}
}
\end{algorithm}

Each process $p_i$ manages a new local variable $ak_i$ (approximate $k$),
initialized to $\sigma_i= \lceil\frac{\Delta_i}{m}\rceil+1$, and whose final 
value will be $K$. The additional behavior of $p_i$ is now as follows.
\begin{itemize}
\vspace{-0.2cm}
\item 
First, when a non-root process informs its parent of its local termination,  
it now broadcasts the message {\sc term}$(parent_i, id_i, ak_i)$ 
(line~\ref{MC-28}'). Hence, the values $ak$ of the leaves will be the 
first to be known by their parent. 
\vspace{-0.2cm}
\item 
 When a process receives a message  {\sc term}$(dest,id, ak)$ 
(line~\ref{MC-30}'), in addition to its previous statements, it updates 
$ak_i$ to  ${\sf max}(ak_i,ak)$ (line N1). It follows that the root
learns the value of $K$ (which is the maximal value $ak$ it receives). When 
it learns it, the root progresses to the local state 5 (line~\ref{MC-34}').

\vspace{-0.2cm}
\item Starting from the root, when a non-leaf process 
is in state 5 and allowed to broadcast (predicate of line N2), 
it broadcasts the message {\sc end}$(id_i,ak)$ and progresses to the 
final state 6  (line N3).

\vspace{-0.2cm}
\item Finally, when a process $p_i$, in state 4, receives a message 
{\sc end}$(sender,ak)$ from its parent, it updates $ak_i$ to 
${\sf max}(ak_i,ak)$ and progresses to state 5 (line N6). 
It will then forward the message {\sc end}$(sender,ak)$ to its children 
if it has some, and in all cases will enter local state 6  (line N3). 
\end{itemize}

After the local state of a process $p_i$ became 6, we have $ak_i=K$. 
Hence, all processes are provided with a round-based programming  
level in  which each process $p_i$ can  C2$m$-freely broadcast  messages
at all the rounds such that $(\CLOCK {\sf~ mod~} ak_i) \in  \colors_i$.

\end{document}